\documentclass{jnmp}
\usepackage{latexsym}
\usepackage{amssymb}
\usepackage{graphicx}

\newcommand{\beq}{\begin{equation}}
\newcommand{\eeq}{\end{equation}}
\newcommand{\bea}{\begin{eqnarray}}
\newcommand{\eea}{\end{eqnarray}}
\newcommand{\barr}[1]{\begin{array}}
\newcommand{\earr}{\end{array}}
\newtheorem{theorem}{Theorem}[section]
\newtheorem{definition}{Definition}[section]

\newcommand{\bdf}{\begin{definition}}
\newcommand{\edf}{\end{definition}}
\newcommand{\bth}{\begin{theorem}}
\newcommand{\enth}{\end{theorem}}
\newcommand{\pd}{\partial}

\newcounter{abc}
\newenvironment{zoznamrom}{\setcounter{abc}{0}\begin{list}{\roman{abc})}
                       {\usecounter{abc}}}{\end{list}} 
\newcounter{zozal}

\newcounter{znum}
\newenvironment{zoznamnum}{\setcounter{znum}{0}\begin{list}{\arabic{znum}.}
                       {\usecounter{znum}}}{\end{list}}

\def\kp{k_{+}} 
\def\km{k_{-}} 
\def\kz{k_{0}}
\def\mR{\mathbb R}
\def\mC{\mathbb C}

\def\zl{\left(}
\def\zr{\right)}
\def\c+{\rlap{\ \raisebox{.2ex}{\scriptsize+}}\supset}

\begin{document}
\renewcommand{\evenhead}{E G Kalnins, Z Thomova and P Winternitz}
\renewcommand{\oddhead}{Separation of variables 
in Hamilton-Jacobi and Schr\H{o}dinger equations}

\thispagestyle{empty}


\copyrightnote{2004}{E G Kalnins, Z Thomova and P Winternitz}

\Name{Subgroup type coordinates and the separation of variables 
in Hamilton-Jacobi and Schr\H{o}dinger equations}

\label{firstpage}

\Author{E.G. Kalnins~$^\dag$ and Z. Thomova~$^\ddag$ and P. Winternitz~$^*$}

\Address{$^\dag$ Department of Mathematics, The University of Waikato,
Private Bag 3105, Hamilton, New Zealand \\
~~E-mail: e.kalnins@waikato.ac.nz\\[10pt]
$^\ddag$ Department of Mathematics and Sciences,
 SUNY Institute of Technology, P.O. Box 3050, Utica, NY 13504 \\
~~E-mail: thomovz@sunyit.edu \\ [10 pt]
$^*$Centre de recherche math{\'e}matiques and D{\'e}partement 
de math{\'e}matiques et de statistique,
Universit{\'e} de Montr{\'e}al, 
Case postale 6128, succ. centre-ville,
Montr{\'e}al, Qu{\'e}bec, H3C 3J7, Canada\\
~~E-mail:wintern@crm.umontreal.ca }


\begin{abstract}
\noindent
Separable coordinate systems are introduced in  complex and 
real four-dimensional flat spaces. We use maximal Abelian subgroups to 
generate  coordinate systems with a maximal number of ignorable variables. 
The results are presented (also graphically)  in terms of subgroup chains. 
Finally the explicit solutions of the  Schr\H{o}dinger equation in the 
separable coordinate systems are computed.
\end{abstract}


\section {Introduction}
The bicentennial of both Carl Gustav Jacob Jacobi and William Rowan Hamilton
 provides us with an excellent opportunity to take a new look at the equation
 associated with both of their names, as well as its twentieth-century 
descendent, the Schr\H{o}dinger equation. The integrability, or superintegrability 
\cite{kkmw,kalmilpog,kalmilwin,KWMP,MSVW,wsuf,wojc}
of these equations, i.e. the existence of $n$, respectively $k$ 
(with $n+1 \leq k \leq 2n-1$) integrals of motion, belongs to the fundamental problems 
concerning any classical, or quantum, Hamiltonian system. Of course much has transpired
since the days of Hamilton and Jacobi. In particular Lie group theory has been created and
its power applied to classical and quantum mechanics.

Among Hamiltonian systems that are integrable a special class consists of those 
that allow the separation of variables in the Hamilton-Jacobi and Schr\H{o}dinger
equations. This occurs typically when the integrals of motion are quadratic polynomials
in the momenta (in classical mechanics) or second-order differential operators 
in quantum mechanics. Superintegrable systems with more than $n$ second-order integrals of
motion are typically multiseparable, i.e. separable in more than one coordinate system.

An extensive literature exists on Lie theory and the separation of variables
\cite{benenti,BKM,eis1,eis2,PogIzW,kalnins,kkmw,KalMil1,KalMil2,kalmilpog,kalmilwin,PogW,V1,
vilenkin,frisw}. 
Experts on the separation 
of variables immediately think of ellipsoidal 
coordinates and degenerate cases thereof. Most ``practitioners"  think of the simplest 
types of coordinates in Euclidean three-space, cartesian, cylindrical and spherical coordinates.
These coordinates have been called ``subgroup type coordinates" because they are related to 
different subgroup chains of the Lie group $G$, the isometry group of the space under 
consideration \cite{kalmilwin,milpatw,PogW,frisw,wsuf}.

The purpose of this article is to analyze further these subgroup type coordinates. They
exist in any space with a nontrivial isometry group. We consider complex and real 
Euclidean 
spaces, as well as pseudo-Euclidean real spaces and their isometry groups $E(n,\mC)$,
$E(n)$ and $E(p,q)$, respectively. We see that a much greater variety of such 
coordinates exists for the $E(n,\mC)$ and $E(p,q)$ groups than for real Euclidean ones.
Moreover some of them have new and interesting properties. The coordinates 
are not necessarily orthogonal ones. The separated ordinary differential
equations are not necessarily of second order; quite often they are first-order equations
and the solutions then involve elementary functions rather than special ones.

Here we consider free motion only, that is there is no potential in the Hamiltonian. 
Once separable coordinates and the corresponding integrals are established, it is an easy
task to add a potential to the Hamiltonian and modify the integrals of motion in such a manner
as to preserve separability \cite{eis1,eis2,kkmw,kalmilwin}. 

From the mathematical point of view this article is an application of a research programme, 
the aim of which is to classify the maximal Abelian subalgebras (MASAs) of all classical 
Lie algebras. In particular in earlier articles \cite{epc,conf,epq} we presented 
a classification of MASAs 
of the $e(n,\mC)$ and $e(p,q)$ algebras. Here we apply this classification to the problem 
at hand. 

The problem we are considering can be posed as follows.
Consider a complex, or real, n-dimensional Riemannian or pseudo-Riemannian 
space S with metric
\beq ds^2=\sum_{i,k=1}^{n} g_{ik}(x) dx^i dx^k 
\eeq
 and an isometry group $G$ of dimension $N \geq n$. We wish to construct all coordinate
 systems that satisfy the following requirements.
\begin{zoznamnum}
\item
They allow the separation of variables in the time-independent free Schr\H{o}dinger equation
\beq \! \! \! \! \! \! \! \! \! \! \! \! \! \! \! \! \! \! \!
H \Psi = E \Psi,  \quad
H = - \frac{1}{2} \frac{1}{\sqrt g} \sum_{i,k=1}^n \frac{\partial}{\partial x^i}
{\sqrt g} g^{ik} \frac{\partial}{\partial x^k}, \quad
g_{ik}g^{kl} = \delta_{il}, \quad g=det(g_{ik}) \label{eq:schr}
\eeq
and also in the Hamilton-Jacobi equation
\beq
\sum_{i,k=1}^n g^{ik} \frac {\partial W} {\partial x_i} 
\frac {\partial W}{\partial x_k} = E. \label{eq:HJ}
\eeq
Thus for the Schr\H{o}dinger equation (\ref{eq:schr}) and Hamilton-Jacobi equation
(\ref{eq:HJ}) we require
\bea
\Psi = \prod_{i=1}^n \Psi_i (x^i, \lambda_1, \ldots , \lambda_n), \label{eq:14} \\
W=\sum_{i=1}^n W_i (x^i, \lambda_1, \ldots , \lambda_n) \label{eq:15}
\eea
respectively, where $\lambda_1, \ldots, \lambda_n$ are constants of separation.

\item
All coordinates $x_i$ are coordinates of subgroup  type. By this we mean that 
a general element of the isometry group $G$ is written as a product of one-dimensional
subgroups $g=g_1 g_2 \ldots g_N$, $N={\rm dim} G$ and the coordinates $x_i$ are generated by 
the action of $G$ on some chosen origin $| o \rangle$
\beq 
|x \rangle =g |o \rangle.
\eeq

\item
Each coordinate system contains a maximal number of ignorable coordinates 
(i.e. variables not figuring
in the metric tensor $g_{ik}(x)$) generated by a maximal Abelian subgroup $G_M \subset G$.
\end{zoznamnum}

The above conditions are satisfied by parametrizing the group element
$g \in G$ as follows
\bea
g=g_M (a_1, \ldots, a_k)h(s_1, \ldots, s_l)g_0(u_1, \ldots , u_m) \label{eq:para} \\
k+l+m = {\rm dim} G, \, \, \, \, \, k+l = {\rm dim} M. \nonumber
\eea

In eq.(\ref{eq:para}) $g_M$ is a maximal Abelian subgroup, $h$ is a product 
of one-parameter subgroups (not necessarily a subgroup itself) and $g_0$ is the isotropy
 group of the origin $ | 0 \rangle$. The group parameters $a_i$ provide ignorable 
variables while the parameters $s_i$ are the ``essential'' variables (that do figure in 
$g_{ik}(x) \equiv g_{ik}(s_1, \ldots, s_l)$).

Not every maximal Abelian subgroup is suitable for this purpose. The requirement is 
that $g_M$, when acting on a generic point in space $S$, should sweep out orbits 
of dimension $k$ (not lower-dimensional ones).

In the process we also solve a more general problem, namely that of constructing all coordinates
of subgroup
type, also those not involving a maximal number of ignorable variables.

We actually work with the Schr\H{o}dinger equation (\ref{eq:schr}) only and 
obtain separable coordinates $(x_1, \ldots , x_n)$ in the sense of eq. (\ref{eq:14}). The additive 
separation (\ref{eq:15}) for the Hamilton-Jacobi equation in the same coordinate system follows
automatically. The separated wave functions (\ref{eq:14}) are eigenfunctions of a complete set
of commuting operators
\beq
\{ X_1, X_2, \ldots , X_n \}. \label{eq:setX}
\eeq
Among them $k$ operators are basis elements of a maximal Abelian subalgebra (MASA) of the
Lie algebra of the isometry group. One operator is the Hamiltonian $H$ and the others are all 
second-order Casimir operators of subgroups in a subgroup chain
\bea
G \supset G_1 \supset G_2 \supset \ldots \supset G_M, \label{eq:chainG} 
\eea
where $G_M$ is a maximal Abelian subgroup of $G$.

More specifically we consider a flat space M with a complex Euclidean, real Euclidean,
 or pseudo-Euclidean
isometry group. We realize its Lie algebra $e(n,\mC)$, $e(n)$, or $e(p,q)$ by matrices
\bea
& E =\left( \begin{array}{cc} X  & \alpha \\
                              0 & 0 
                              \end{array} \right),
\qquad X \in F^{n \times n}, \qquad \alpha \in F^{n \times 1} \label{eq:E}, & \\
& XK+KX^{T}=0, \qquad K=K^T, \qquad {\rm det} K \neq 0, \nonumber &
\eea
where we have $F=\mR$, or $F=\mC$. For $F=\mR$ the matrix $K$ determining the metric has
 a signature
\beq
{\rm sgn} K = (p,q), \qquad p \geq q \geq 0
\eeq
(i.e. $p$ positive and $q$ negative eigenvalues).

We use several different choices of the matrix $K$. Different choices of $K$ and 
$X$ are related by the transformation
\beq
G K_1 G^T = K_2, \qquad G X_1 G^{-1} = X_2, \qquad G \subset GL(N,F).
\eeq
Coordinates in the space $S$ have the form
\beq
| y \rangle = \left( \begin{array}{c} | x \rangle  \\
                                           1 
                              \end{array} \right),
\quad x \in F^{n \times 1}, \quad | 0 \rangle =  \left( \begin{array}{c}| o \rangle  \\
                                           1 
                              \end{array} \right),
\eeq
where $ | o \rangle$ is chosen to be the  origin of the space $M$. The isotropy group
of the origin is $G_0 \sim O(n,\mC)$ for $F=\mC$, $G_0 \sim O(p,q)$, 
$p+q=n$ for $F = \mR$.

We give some decomposition theorems in Section 2 for arbitrary $n$ and then 
concentrate on
the case $n=4$.
This is the most interesting case from the point of view of physical applications, mainly
in the context of special and general relativity \cite{BKM,VilSm,frisw}, but also in that 
of integrable and 
superintegrable systems \cite{KWMP,MSVW,ORW1}. Separation of variables also plays a role in the 
study of Huygens' principle \cite{berest}, where dimension 4 is again the one of physical 
importance.

From the mathematical point of view ${\rm dim} M =4$ is sufficiently simple so that it can 
be treated in  a complete and detailed manner. On the other hand it is rich enough to 
demonstrate most
of the phenomena that occur for any value of $n$.



\section{Decompositions of spaces and algebras}
\setcounter{equation}{0}
\subsection{Chains of subgroups and decompositions of $M(n)$}

The role of subgroup chains in the study of coordinate separation has 
been emphasized in many articles \cite{PogIzW,milpatw,PogW,VKS}. For real Euclidean 
groups $E(n)$ and their Lie algebras $e(n)$ the situation is quite simple.
 Only two types of maximal subgroups exist, namely the following

\begin{eqnarray} 
E(n)& \supset & E(n_1) \otimes E(n_2), \quad  n_1+n_2 =n, \quad  n_1 \geq n_2 \geq 1  \label {eq:En1} \\
E(n)& \supset & O(n), \qquad \qquad \qquad  n \geq 2. \label{eq:En2} 
\end{eqnarray}

Maximal subgroups of $O(n)$ can be embedded in the defining representation
 of $O(n)$ reducibly or irreducibly. The reducibly embedded ones leave 
a vector subspace of the Euclidean space $M(n)$ invariant. The corresponding
 maximal subgroups are one of the following:
\bea
O(n)& \supset & O(n_1) \otimes O(n_2),\quad n_1+n_2 = n, \quad  n_1 \geq n_2 \geq 2 
\label{eq:On1} \\
O(n) & \supset & O(n-1),  \qquad  \qquad n \geq 3. \label{eq:On2}
\eea
The subgroup link (\ref{eq:En1}) leads to the decomposition of the
Euclidean space $M(n)$ into the direct sum of two Euclidean subspaces
\beq
M(n) = M(n_1) \oplus M(n_2), \quad n_1+n_2=n, \quad n_1 \geq n_2 \geq n.
\eeq
Separable coordinates can then be introduced separately on $M(n_1)$ and $M(n_2)$; 
a lower dimensional task. The subgroup chain (\ref{eq:En2}) leads to the embedding 
of a sphere $S_{n-1}$ into $M(n)$. The links  (\ref{eq:On1}) and  (\ref{eq:On2}) lead
to various types of spherical and polyspherical coordinates on this sphere 
\cite{PogIzW,V1,vilenkin}.

Irreducibly embedded subgroups of $O(n)$, like $U(n) \subset O(2n)$, or $G_2 \subset O(7)$,
have not been used to generate separable coordinates.

For complex Euclidean groups $E(n,\mC)$ and pseudo-Euclidean groups $E(p,q)$  the
first subgroup links are essentially the same as (\ref{eq:En1}) and (\ref{eq:En2})
(mutatis mutandis), e.g. for $E(p,q)$ we have two possibilities:

\bea
E(p,q) & \supset & E(p_1,q_1) \otimes E(p_2,q_2), \quad p_1+p_2 = p, \qquad  q_1+q_2=q 
\label{eq:Epq1} \\
E(p,q) & \supset & O(p,q), \qquad  \qquad \qquad \quad n=p+q \geq 2. \label{eq:Epq2}
\eea
However, the subgroup links of the type (\ref{eq:On1}) and (\ref{eq:On2}) are not 
the only ones for $O(n,\mC)$, or $O(p,q)$ with $p \geq q \geq 1$. In particular the 
possibilities for 
$O(p,1)$ were discussed in Ref. \cite{PogW} and are 
\bea
O(p,1) & \supset & O(p), \qquad \qquad \qquad O(p,1) \supset O(p-1,1) \\
 O(p,1)& \supset & O(p_1,1) \otimes O(p_2), \qquad p_1+p_2 =p, \quad p_1 \geq 1, p_2 \geq 2 
\eea
and also
\beq
O(p,1) \supset E(p-1).
\eeq
For $O(p,q)$, $p\geq q \geq 2$, the possibilities are even richer as we 
see below in the case of $O(2,2)$.
 
In any case we are not interested in subgroup links of the type 
(\ref{eq:En1}), or (\ref{eq:Epq1}) since they lead to a decomposition 
of the considered space M and we assume that the problem is already solved 
in lower dimensions. In other words we are only interested in ``indecomposable'' 
coordinate systems in the space M.


\subsection{Decomposition of MASAs}
Maximal Abelian subalgebras (MASAs) of Euclidean Lie algebras can  contain
generators of translations. In the real case, $e(n)$, these translations can only be
space-like and their presence in a subgroup chain leads to the decomposition of the
space $M(n)$. For $E(n, \mC)$ there can be up to $[ \frac{n}{2}]$ lightlike 
(isotropic) translations
 in a MASA and
for $e(p,q)$, $p \geq q \geq 1$, up to $q$ lightlike translations. These do not 
lead to a decomposition of $M(n)$, or $M(p,q)$ respectively and must hence be considered. 
The algebras $o(n,\mC)$ allow three types of MASAs \cite{verc}:
\begin{zoznamnum}
\item
Orthogonally decomposable (OD) MASAs
\item
MASAs that are decomposable but not orthogonally (OID but D)
\item
Indecomposable MASAs (OID and ID).
\end{zoznamnum}
The MASAs of $o(n,\mC)$ have been classified elsewhere \cite{verc}. The algebras $o(p,q)$
have up to six types of MASAs, depending on their decomposability properties
over the real numbers, and their behavior under complexification \cite{verop}. All 
types of MASAs occur in $e(4, \mC)$ and $e(2,2)$, respectively.


\section{Separable coordinates in the complex space $M(4,\mC)$}
\setcounter{equation}{0}

\subsection{MASAs of $e(n,\mC)$.}
The MASAs of $e(n,\mC)$ have been classified into conjugacy classes under the action 
of the group $E(n,\mC)$ in an earlier publication \cite{epc}. Each class of MASAs is 
represented by one ``canonical'' MASA. In any basis a MASA of $e(n,\mC)$ contains
$k_0+k_1$ mutually orthogonal translations, $k_0$ of them isotropic, $k_1$ anisotropic. 
We have
\beq
0 \leq k_0 \leq \left[ \frac{n}{2} \right], \qquad 0 \leq k_1 \leq n, \qquad 0 \leq k_0+k_1 \leq n.
\eeq

We are mainly interested in MASAs with $k_1=0$ since the presence of anisotropic 
translations leads to a decomposition of the space into 
$M(n,\mC) \equiv M(k_1,\mC) \oplus M(n-k_1,\mC)$ and hence to decomposable coordinate systems.

We now consider the four-dimensional space $M(4,\mC)$. We denote the MASAs 
of $e(4,\mC)$ as $M_{4,j}(k_0)$. We run through all inequivalent MASAs and construct 
the separable coordinates. In each case we represent the MASA by a matrix $X \in \mC^{5 \times 5}$
and also the corresponding metric $K=K^T \in \mC^{4 \times 4}$, ${\rm det} K \neq 0$ 
(see eq. (\ref{eq:E})). The coordinates are generated by a group action as in eq. (\ref{eq:para}). 
More specifically we write
\bea
| y \rangle = \left( \begin{array}{c} | x \rangle \\ 1 \end{array} \right)=
g \left( \begin{array}{c} | o \rangle \\ 1 \end{array} \right), \qquad
| x \rangle =  \left( \begin{array}{c} x_1\\ x_2 \\ x_3 \\ x_4 \end{array} \right), \quad
| o \rangle =  \left( \begin{array}{c} 0\\ 0 \\ 0 \\ 0 \end{array} \right).
\eea

The group action is specified by giving $g = g_1 g_2 g_3 g_4$, where each $g_i$
 is a one-dimensional subgroup of $E(4,\mC)$. The MASAs are either two- or three-dimensional 
(for indecomposable coordinate systems). They generate $g_1$ and $g_2$ 
for two-dimensional MASAs and $g_1, g_2$ and $g_3$ for three-dimensional ones.

We also give the complete sets of commuting operators (\ref{eq:setX}) related to each MASA. 
For three-dimensional MASAs the set consists of $\{\Box_4, X_1, X_2, X_4\}$, 
where $\Box_4$ is the Laplace operator and $\{X_1, X_2, X_3\}$ generates the MASA. For
two-dimensional MASAs the set consists of $\Box_4, X_1, X_2$ and $ Y$, where
$Y$ is a second-order Casimir operator of a subalgebra $L$ satisfying
\beq
\{X_1,X_2\} \subset L \subset e(4,\mC).
\eeq
In each case we give $L$ and its Casimir operator.

The notation $E_{ik}$ denotes the matrices satisfying $(E_{ik})_{ab}=\delta_{ia} \delta_{kb}$,
always in a basis corresponding to the metric $K$.


\subsection{MASAs of $e(4,\mC)$ with $k_0=0$}

There are two MASAs with $k_0=0$. 
\begin{zoznamnum}
\item The Cartan subalgebra $M_{4,1}(0) \sim o(2,\mC) \oplus o(2,\mC)$
\bea M_{4,1}(0) =\left( \begin{array}{ccccc} 0 & -a  & 0 & 0 & 0\\
                                        a & 0 & 0 & 0 & 0 \\
                                   0 & 0 & 0  & -b & 0   \\
                                         0 & 0 & b & 0 & 0 \\
					 0 & 0 & 0 & 0 & 0 \\
                  \end{array} \right), 
\quad
K = \left( \begin{array}{cc} I_2 & 0 \\
                             0 & I_2 \\
          \end{array} \right).   \label{eq:MV41}
\eea
The coordinates on the complex Euclidean space $M_4(\mC)$ are generated 
by the following group action on the origin: 
\bea
|y \rangle =e^{-a (E_{12}-E_{21})}e^{-b (E_{34}-E_{43})}
e^{-c (E_{13}-E_{31})}e^{r E_{15}} | 0 \rangle. \label{eq:M41-00}
\eea
The variables then are:
\bea \begin{array}{lll}
x_1  = r \cos c \cos a &  & x_3  =   r \sin c \cos b \\
x_2 =  r \cos c \sin a & &  x_4 =   r \sin c \sin b. 
\end{array} \label{eq:V4100} \eea
Here $r,c,a$ and $b$ are all complex. For $r$ constant we have 
a complex sphere on which $a$ and $b$ provide cylindrical coordinates.
The subgroup chain that provides these coordinates and the complete set of
commuting operators is
\beq
E(4, \mC) \supset O(4,\mC) \supset O(2,\mC) \otimes O(2,\mC).
\eeq

The Laplace operator  $\Box_{4\mC}$ in this case is
\bea
\Box_{4\mC} = \frac{\partial^2}{\partial r^2} + \frac{3}{r} \frac{\partial}{\partial r} 
+ \frac{1}{r^2} \bigtriangleup_{LB}, \label{eq:318}
\eea
where
\bea
\bigtriangleup_{LB} =\frac{\partial^2}{\partial c^2} + 2 \cot 2 c \frac{\partial}{\partial c}+
\frac{1}{\cos^2 c} \frac{\partial^2}{\partial a^2} + 
\frac{1}{\sin^2 c} \frac{\partial^2}{\partial b^2}
\eea
is the Laplace-Beltrami operator on the sphere $S_4(\mC)$ (the Casimir operator 
of $O(4, \mC)$ that does not vanish on the sphere). The complete set of commuting operators 
in this case consists of
\bea
\{\Box_{4\mC},\, \bigtriangleup_{LB},\, L_{12}=E_{12}-E_{21},\,  L_{34}=E_{34}-E_{43}\}.
\eea

The one parameter subgroups $L_{12}$, $L_{34}$ and $L_{13}=E_{13}-E_{31}$ 
figuring in the chain (\ref{eq:M41-00})
generate all of $O(4,\mC)$.
\item 
The orthogonally indecomposable, but decomposable (OID but D) MASA of $o(4,\mC)$
\bea M_{4,2}(0) =\left( \begin{array}{ccccc} a & b  & 0   & 0  & 0 \\
                                             0 & a  & 0   & 0  & 0 \\
                                             0 & 0  & -a  & 0  & 0   \\
                                             0 & 0  & -b  & -a & 0 \\
					     0 & 0  & 0   & 0  & 0 \\
                  \end{array} \right), 
\quad
 K=\left( \begin{array}{cc} 0 & I_2 \\
                             I_2 & 0 \\
          \end{array} \right). \label{eq:M420}        
\eea
The coordinates on $M_4(\mC)$ are generated by the following group action:
\bea \! \! \! \! \! \! \! \! \! \! \! \! \! \! \! \! \!
|y \rangle = e^{a(E_{11}+E_{22}-E_{33}-E_{44})}e^{b(E_{12}-E_{43})} 
e^{c(E_{11}-E_{22}-E_{33}+E_{44})}
e^{r(E_{15}+E_{25}+E_{35}+E_{45})}| 0 \rangle \label{eq:M42-00gch}
\eea

We get
\bea \begin{array}{lll}
x_1  =  \frac{1}{2}r e^a (e^c + b e^{-c})& &x_3  =  \frac{1}{2} r e^{-a} e^{-c}   \\
x_2  = \frac{1}{2} r e^a e^{-c} & & x_4 = \frac{1}{2} r e^{-a} (e^c - b e^{-c}).
\end{array} \label{eq:V4200} \eea

The subgroup chain is 
\bea
E(4,\mC) \supset O(4,\mC) \supset {\rm exp} (M_{4,2}(0))
\eea
and the complete set of commuting operators is
\bea
\{\Box_{4 \mC},\, \bigtriangleup_{LB},\, X_1, \, X_2 \}, \label{eq:316}
\eea
where $\Box_{4\mC}$ is the operator in (\ref{eq:318}) and 
$\bigtriangleup_{LB}$ is 

\bea
\bigtriangleup_{LB}= - \frac {\partial^2}{\partial c^2}
+ 2 \frac {\partial}{\partial c}
+ 4 e^{4c} \frac {\partial}{\partial b^2} 
- 4 e^{2c} \frac{\partial^2 }{\partial a \partial b}. \label{eq:317}
\eea
The first three subgroups in (\ref{eq:M42-00gch}) are all contained in a subgroup 
 $GL(2,\mC) \subset O(4,\mC)$. However, this subgroup does not have a Casimir operator
(though it leaves that of $O(4,\mC)$ invariant).

The coordinates (\ref{eq:V4200}) lead to a nonorthogonal separation of variables in 
$\bigtriangleup_{LB}$ and hence in $\Box_{4 \mC}$.
\end{zoznamnum}


\subsection{MASAs of $e(4,\mC)$ with $k_0=1$}

All MASAs of $e(4,\mC)$ containing one isotropic translation can be written in the form
\cite{epc,epq}
\bea
X=\left( \begin{array}{cccc} 0 & \alpha  & 0  & z \\
                               0 & 0  & -K_0 \alpha^T & BK_0\alpha^T \\
                                0 & 0 & 0 & 0   \\
                                0 & 0 & 0 & 0  
                  \end{array} \right), \quad
K=\left( \begin{array}{ccc} 0 & 0 & 1\\
                            0 & K_0 & 0  \\
                            1 & 0 & 0 
                           \end{array} \right), \label{eq:313} \\
\begin{array}{l}
\alpha = (a_1, a_2), \quad K_0=K_0^T \in \mC^{2 \times 2}, \quad {\rm det} K_0 \neq 0, \nonumber \\
BK_0=K_0 B^T, \qquad  B \in \mC^{2 \times 2}. \nonumber
\end{array}
\eea
The pair of matrices $(B,K_0)$ can be transformed into one of the following standard forms 
\cite{djok}
\bea
B_1&=&\left( \begin{array}{cc} 0 & 0   \\
                               0 & 0  
                  \end{array} \right), \quad
K_0=I_2 \\
B_2&=&\begin{array}{ll}
\left( \begin{array}{cc} 1 & 0   \\
                          0 & \beta  
       \end{array} \right), &
\beta = |\beta| e^{i \phi}, \quad \left\{ \begin{array}{ll}
0 \leq |\beta|<1, & \quad 0 \leq \phi < 2 \pi \\
|\beta| = 1, & \quad  0 \leq \phi < \pi \end{array} \right. \\ 
& K_0=I_2 \end{array} \label{eq:320} \\
B_3&=&\left( \begin{array}{cc} \kappa & 0   \\
                               1 & \kappa  
                  \end{array} \right), \quad
\kappa =\left\{ \begin{array}{c} 0 \\ 1
                  \end{array} \right. , \quad
K_0= \left( \begin{array}{cc} 0 & 1   \\
                               1 & 0  
                  \end{array} \right).
\eea
We mention that the MASAs (\ref{eq:313}) are maximal Abelian nilpotent subalgebras 
of $e(4,\mC)$ (MANS) \cite{suprtysh,decomp}. This implies that they are represented 
by nilpotent matrices in any finite dimensional representation.

The corresponding MASAs are all three-dimensional and directly provide three 
commuting operators and three ignorable variables. We continue our list of MASAs and coordinates:

\begin{zoznamnum}
\item
\bea M_{4,3}(1) =\left( \begin{array}{ccccc} 0 & -a_1  & -a_2 & 0 & z \\
                                        0 & 0 & 0 & a_1 & 0 \\
                                   0 & 0 & 0 & a_2 & 0   \\
                                         0 & 0 & 0 & 0 & 0  \\
                                    0 & 0 & 0 & 0 & 0 \\
                  \end{array} \right), \quad
K =\left( \begin{array}{ccc}   0 & 0 & 1 \\
			       0 & I_2 & 0 \\
			       1 & 0 & 0 
           \end{array} \right) 
\eea
\bea
| y \rangle =e^{a_1(-E_{12}+E_{24})} e^{a_2(-E_{13}+E_{34})}e^{z E_{15}}e^{r E_{45}} |0 \rangle
\eea

\bea \begin{array}{llll}
x_1  =  z -\frac{1}{2} r (a_1^2+a_2^2) & & & x_3  =  r a_2 \\
x_2  =  r a_1 &  & & x_4  =  r .  
\end{array} \label{eq:V431} \eea

The $M(4,\mC)$ Laplace operator with these new variables is
\bea
\Box_{4C} =2 \frac{\partial^2}{\partial z \partial r} 
+\frac{2}{r} \frac{\partial}{\partial z} 
+ \frac{1}{r^2}\left( \frac{\partial^2}{\partial a_1^2}+ \frac{\partial^2}{\partial a_2^2}\right).
\label{eq:L325}
\eea

\item
\bea 
M_{4,4}(1) =\left( \begin{array}{ccccc} 0 & -a_1  & -a_2 & 0 & z\\
                                        0 & 0 & 0 & a_1 & a_1  \\
                                   0 & 0 & 0 & a_2 & a_2 \beta  \\
                                    0 & 0 & 0 & 0 & 0  \\
                                    0 & 0 & 0 & 0 & 0 
                  \end{array} \right), \quad
K =\left( \begin{array}{ccc}  0 & 0 & 1 \\
			      0 & I_2 & 0 \\
			      1 &  0 & 0 
	    \end{array} \right)  
\eea

\bea
|y \rangle =e^{a_1(-E_{12}+E_{24}+E_{25})} e^{a_2(-E_{13}+E_{34}+\beta E_{35})}
e^{z E_{15}}e^{r E_{45}} |0 \rangle 
\eea

\bea \begin{array}{l} \begin{array}{llll}
x_1  =  z - \frac{1}{2}r(a_1^2 +a_2^2)-\frac{1}{2}( a_1^2+\beta a_2^2) & & &
x_3  =  (r+\beta) a_2\\
x_2  =  (r+1) a_1 & & &
x_4=r \end{array}  \\
\beta = |\beta| e^{i \phi}, \quad \left\{ \begin{array}{ll}
 0 \leq |\beta|<1, & \quad 0 \leq \phi < 2 \pi \\
|\beta| = 1 & \quad  0 \leq \phi < \pi \end{array} \right.\end{array} \label{eq:V441}
\eea

The Laplace operator is
\bea
\Box_{4C} = 2 \frac{\partial^2}{\partial r \partial z}+
\frac{2r +\beta +1}{(r+1)(r+\beta)} \frac{\partial}{\partial z}
+ \frac{1}{(r+1)^2}\frac{\partial^2}{\partial a_1^2} 
+\frac{1}{(r+\beta)^2}
\frac{\partial^2}{\partial a_2^2}.
\label{eq:L329}
\eea

\item
\bea 
M_{4,5}(1) =\left( \begin{array}{ccccc} 0 & -a_1  & -a_2 & 0 & z\\
                                        0 & 0 & 0 & a_2 & \kappa a_2 \\
                                   0 & 0 & 0 & a_1 & \kappa a_1+a_2  \\
                                         0 & 0 & 0 & 0 & 0  \\
                                    0 & 0 & 0 & 0 & 0 \\
                  \end{array} \right), \, \,
K =\left( \begin{array}{cccc} 0 &  0 & 0 & 1 \\
				0 &  0 & 1 & 0 \\
				0 &  1 & 0 & 0 \\
				1 &  0 & 0 & 0 
                                   \end{array} \right) 
\eea
\bea
|y \rangle =e^{a_1(-E_{12}+E_{34}+\kappa E_{35})} e^{a_2(-E_{13}+E_{24}+ E_{35}+\kappa E_{25})}
e^{zE_{15}}e^{rE_{45}}| 0 \rangle
\eea
\bea \begin{array}{lllcr}
x_1  =  z-(r+\kappa)a_1 a_2 - \frac{1}{2} a_2^2&& x_3 =  (r + \kappa)a_1 +a_2 && \\
x_2  =  (r+ \kappa) a_2   && x_4  =  r && \kappa = 0 \quad  {\rm or } \quad 1.
\end{array} \label{eq:V451}
\eea

The Laplace operator in these coordinates is
\bea
\Box_{4C} = 2 \frac{\partial^2}{\partial r \partial z}
+ \frac{2}{(r+\kappa)} \frac{\partial}{\partial z}
-\frac{2}{(r+\kappa)^3}\frac{\partial^2}{\partial a_1^2}
+\frac{2}{(r+\kappa)^2}\frac{\partial^2}{\partial a_1 \partial a_2}.
\label{eq:L333}
\eea
\end{zoznamnum}

\subsection{ MASAs of $e(4, \mC)$ with $k_0 = 2$}

There are two mutually inequivalent MASAs of $e(4, \mC)$ with 2 isotropic translations
 \cite{epc,epq}.
Both of them are three-dimensional. One of them has three dimensional orbits in $M(4,\mC)$ and 
the other 
only two-dimensional ones. We consider the two of them separately.

\bea M_{4,6}(2) =\left( \begin{array}{ccccc} 0 & 0  & 0 & z & a_1 \\
                                          0 & 0 & -z & 0 & a_2 \\
                                          0 & 0 & 0 & 0 & 0   \\
                                          0 & 0 & 0 & 0 & z  \\
                                          0 & 0 & 0 & 0 & 0 
                  \end{array} \right), \quad
K=\left( \begin{array}{cc} 0 & I_2 \\
                         I_2 & 0    \\
        \end{array} \right).
\eea
Coordinates are generated by the action:
\bea
| y \rangle =e^{z(E_{14}-E_{23}+E_{45})}e^{a_1 E_{15}}e^{a_2 E_{25}}e^{r E_{35}} | 0 \rangle
\eea
\bea \begin{array}{llll}
x_1 =  a_1+{1 \over 2}z^2 &&& x_3 =  r  \\
x_2 = a_2 -r z &&& x_4  =  z.
\end{array} \label{eq:V462}
\eea
The Laplace operator is
\bea
\Box_{4C} = 2 \left( \frac{\partial ^2}{\partial a_1 \partial r}+r \frac{\partial^2}
{\partial a_2^2}+
\frac {\partial^2}{\partial a_2 \partial z} \right). \label{eq:L337}
\eea


The other MASA is represented by
\bea M_{4,7}(2) =\left( \begin{array}{ccccc} 0 & 0  & 0 & z & a_1 \\
                                             0 & 0 & -z & 0 & a_2 \\
                                             0 & 0 & 0 & 0 & 0   \\
                                             0 & 0 & 0 & 0 & 0  \\
                                             0 & 0 & 0 & 0 & 0 
                  \end{array} \right), \quad
K=\left( \begin{array}{cc} 0 & I_2 \\
                           I_2 & 0 \\
         \end{array} \right). \label{eq:338}
\eea
The orbits that the 3-parameter subgroup sweeps out are only 2-dimensional. 
This is reflected in the fact that the subalgebra generators when acting on 
$M(4,\mC)$ are represented, 
in the considered metric, by the commuting operators
\beq
P_1 = \partial_{s_1}, \quad P_2 = \partial_{s_2}, \quad B = s_4 \partial_{s_1} - s_3 \partial_{s_2}.
\eeq
These are linearly connected, i.e. they span a two-dimensional subspace of the tangent space
rather than a three-dimensional one.
The operator B can hence not be straightened, i.e. $s_1 = a_1$ and $s_2=a_2$ 
are already ignorable variables. Hence $B$ must be dropped from the MASA. 
The algebra ${P_1,P_2}$ itself generates coordinates, equivalent to Cartesian ones in $M(4,\mC)$ 
in which all four translations are simultaneously diagonalized and the space $M(4,\mC)$ 
is decomposed.
\bea \begin{array}{llll}
x_1  =  a_1 &&& x_3  =  s_3 \\
x_2  =  a_2 &&& x_4  =  s_4. \\
\end{array}
\eea
The Laplace operator is
\bea
\Box_{4C} = 2 \left( \frac{\partial ^2}{\partial a_1 \partial s_3}+
\frac {\partial^2}{\partial a_2 \partial s_4} \right).
\label{eq:L341}
\eea

\subsection{Decompositions of $M(4,\mC)$}
Possible decompositions of $M(4, \mC)$ are
\bea \begin{array}{ll}
M(4,\mC)= M(3,\mC) \oplus M(1, \mC), & \qquad M(4,\mC)= 2 M(2, \mC)  \\
M(4,\mC) = M(2, \mC) \oplus 2 M(1,\mC),& \qquad M(4,\mC)=4 M(1, \mC).
\end{array} \label{eq:342} \eea
In order to obtain a complete list of subgroup type coordinates on $M(4,\mC)$ we
must also take these decompositions into account and introduce 
indecomposable coordinate systems on them. Each $M(1, \mC)$ space
corresponds to a Cartesian coordinate. Each $M(2,\mC)$ corresponds to
complex polar coordinates.

Two types of indecomposable coordinate systems exist on $M(3,\mC)$,
corresponding to two different MASAs, both of them two-dimensional.
The two MASAs of $e(3,\mC)$ can be written as

\bea X =\left( \begin{array}{cccc} 0 & a & 0 & z \\
                                   0 & 0 & -a & 0 \\
                                   0 & 0 & 0 & -\kappa a   \\
                                   0 & 0 & 0 & 0 
                  \end{array} \right), \quad
K =\left( \begin{array}{ccc} 0 & 0 & 1 \\
                                0 & 1 & 0  \\
                                1 & 0 & 0  
                                \end{array} \right) 
\eea
with $\kappa = 0$ for the first and $\kappa =1$ for the second. We will consider 
the two cases separately.

Case I: $\kappa=0$\\
The coordinates are induced by the action
\bea
| y \rangle =e^{a(E_{12}-E_{23})}e^{z E_{14}}e^{r E_{34}} |0 \rangle
\eea
with $|0 \rangle = ( 0 \, 0 \, 0 \, 1)^T$. The coordinates on $M(3,\mC)$ are
\bea
x_1  =  z - \frac{1}{2} r a^2, \qquad
x_2  =  -ar, \qquad
x_3  =  r.  \label{eqM3C_kapa0} 
\eea
The Laplace operator on $M(3,\mC)$ is
\bea
\Box_{3C}=2 \frac{\pd^2}{\pd r \pd z}+\frac{1}{r^2} \frac{\pd^2}{\pd a^2}
+\frac{1}{r}\frac{\pd}{\pd z} \label{eq:346_kapa0}.
\eea
We mention that these coordinates
are conformally equivalent to Cartesian ones on $M(3,\mC)$ \cite{berest,conf}.

Case II: $\kappa=1$\\
The coordinates are induced by the following action
\bea
| y \rangle =e^{a(E_{12}-E_{23}-E_{34})}e^{z E_{14}}e^{r E_{24}} |0 \rangle
\eea
with $|0 \rangle = ( 0 \, 0 \, 0 \, 1)^T$. The coordinates on $M(3,\mC)$ are
\bea
x_1  =  z +ar +  \frac{1}{6} a^3, \qquad
x_2  =  r+\frac{1}{2} a^2, \qquad
x_3  =  -a. \label{eqM3C_kapa1}
\eea
The Laplace operator on $M(3,\mC)$ is
\bea
\Box_{3C}= \frac{\pd^2}{\pd r^2} -2 \frac{\pd^2}{\pd a\pd z}
+2r \frac{\pd^2}{\pd z^2}. \label{eq:346_kapa1}
\eea



\subsection{Subgroup type coordinates with fewer ignorable variables}

For completeness we consider indecomposable subgroup type coordinates 
that are not related to maximal Abelian subgroups of $E(n,\mC)$. We still parametrize
a group element as in eq. (\ref{eq:para}), but replace the subgroup $g_M$ by $g_A$, where
$g_A \subset g_M$, i.e. $g_A$ is Abelian, but not maximal. Indecomposability requires
that the subgroup chain (\ref{eq:chainG}) start out as $E(n,\mC) \supset O(n,\mC)$. After that
links of the type $O(n,\mC) \supset O(n-1,\mC)$, $O(n,\mC) \supset O(n_1,\mC) \otimes O(n_2,\mC)$ 
and $O(n,\mC) \supset E(n-2, \mC)$ are allowed.

In the metric $K=I_n$ the $e(n-2)$ subalgebra of $o(n,\mC)$ is represented by the matrices
\bea
X =\left( \begin{array}{ccc} 0 & \alpha & 0  \\
                         -\alpha^T & A & -i \alpha^T  \\
                                   0 & i \alpha & 0                                   
                  \end{array} \right), \quad \alpha \in \mC^{1 \times (n-2)},\quad K=I_n,\\
A=-A^T \in \mC^{(n-2) \times (n-2)}.
\eea

For $M(n,\mC)$ we use the metric $K=I_n$ and induce the coordinates by the action
\bea
| y \rangle = g_1 g_2 \ldots g_n |0 \rangle
\eea
with
\bea
g_n=e^{r E_{1, n+1}}, \quad \qquad g_{n-1} = e^{c(-E_{1n}+E_{n1})}. \label{eq:349}
\eea
The other one-parameter subgroups depend on the subgroup chain considered. 
The action of $g_n$ takes us onto
a complex sphere of radius $r$. The one-parameter subgroups $g_1, \ldots g_{n-1}$ 
introduce coordinates on this sphere.

We now specialize to the case of $M(4,\mC)$. We use the metric $K=I_4$. 
A basis for the algebra $o(4,\mC)$ can be chosen to be
\beq
L_{ik}=-E_{ik}+E_{ki}, \qquad 1 \leq i \leq k \leq4.
\eeq
An alternative basis, to be used when the $e(2, \mC)$ subalgebra is important, is
\bea \begin{array}{lll}
L_{23},&\quad  X_1 = L_{12}- i L_{24},& \quad X_2=L_{13}-iL_{34}  \\
L_{14},& \quad Y_1 = L_{12}+iL_{24},& \quad Y_2=L_{13}+iL_{34}. \end{array}
\eea

The complete set of commuting operators is
\beq
\{ \Box_{4 \mC}, \bigtriangleup_{LB}(4), R_2, R_1\},
\eeq
where $R_2$ and $R_1$ must be specified in each case and $R_1$ is an element of $o(4,\mC)$.
Similarly $g_4$ and $g_3$ are as in eq. (\ref{eq:349}), but the one-parameter 
subgroups $g_2$ and $g_1$ are specified in each case.

Four subgroup chains and four types of separable coordinates occur.
\begin{zoznamnum}
\item
$E(4,\mC) \supset O(4, \mC) \supset O(3,\mC) \supset O(2,\mC)$ \\
We take $g_1 = e^{a L_{12}}$, $g_2=e^{bL_{13}}$ and we obtain complex spherical coordinates
\bea \begin{array}{llll}
x_1  = r \cos c \cos b \cos a &&& x_3  =  r \cos c \sin b  \\
x_2  =  r \cos c \cos b \sin a &&& x_4  =  r \sin c.
\end{array}  \label{eq:4C1}  
\eea
We have:
\beq
R_2 = \bigtriangleup_{LB}(3), \qquad \qquad R_1 = L_{12},
\eeq
\beq
\Box_{4C}=\frac{\partial^2}{\partial r^2}+\frac{3}{r}\frac{\partial}{\partial r}
+\frac{1}{r^2}\bigtriangleup_{LB}(4), \label{eq:359}
\eeq
\beq
\bigtriangleup_{LB}(4)=
\frac{\partial^2}{\partial c^2}
 - 2 \tan  c \frac{\partial}{\partial c}
+\frac{1}{\cos^2 c} \frac{\partial^2}{\partial b^2}
+ \frac{1}{\cos^2 c \cos^2 b} \frac{\partial^2}{\partial a^2}
-\frac{\tan b}{\cos^2 c} \frac{\partial}{\partial b}.
\eeq
\item
$E(4,\mC) \supset O(4, \mC) \supset O(3, \mC) \supset E(1, \mC)$ \\
We choose $g_1=e^{a X_1}$ and $g_2=e^{b L_{13}}$. The coordinates are
\bea \begin{array}{llll}
x_1  =  r \cos c (\cos b - \frac{1}{2} a^2 e^{ib}) &&& 
x_3  =  r \cos c (\sin b -\frac{1}{2} ia^2 e^{ib})  \\
x_2  =  r \cos c \, a \, e^{ib} &&& x_4  =  r \sin c 
\end{array}  \label{eq:4C2} 
\eea
and we have 
\beq
R_2 = \bigtriangleup_{LB}(3), \qquad \qquad R_1=X_1.
\eeq
The d'Alambertian $\Box_{4C}$ is as in eq. (\ref{eq:359}) with
\beq
\bigtriangleup_{LB}(4)=\frac{\partial^2}{\partial c^2}-2 \tan c \frac{\partial}{\partial c}
+\frac{1}{\cos^2 c}\bigtriangleup_{LB}(3), 
\eeq
\beq
\bigtriangleup_{LB}(3)=\frac{\partial^2}{\partial b^2} + i \frac{\partial}{\partial b} +
e^{-2ib}\frac{\partial^2}{\partial a^2}.
\eeq
\item
$E(4,\mC) \supset O(4,\mC) \supset E(2,\mC) \supset O(2,\mC)$ \\
We choose $g_1 = e^{aL_{23}}$ and $g_2=e^{bX_1}$. The separable coordinates are
\bea \begin{array}{llll}
x_1 =  r (\cos c - \frac{1}{2} b^2 e^{ic})&&& x_3  =  r b e^{ic} \sin a  \\
x_2  =  r b e^{ic} \cos a &&& x_4 =  r (\sin c -\frac{1}{2} ib^2 e^{ic}).\
\end{array}  \label{eq:4C3} 
\eea
We have
\beq
R_2 = \bigtriangleup(2), \qquad \qquad R_1 = L_{23}.
\eeq
Again $\Box_{4C}$ is as in (\ref{eq:359}) with
\beq
\bigtriangleup_{LB}(4)=\frac{\partial^2}{\partial c^2} + 2i \frac{\partial}{\partial c}
+e^{-2ic}\bigtriangleup(2),
\eeq
where
\beq
\bigtriangleup(2)=\frac{\partial^2}{\partial b^2} +\frac{1}{b} \frac{\partial}{\partial b}
+\frac{1}{b^2} \frac{\partial^2}{\partial a^2}.
\eeq

\item
Chain $E(4,\mC) \supset O(4,\mC) \supset E(2, \mC) \supset E(1,\mC) \otimes E(1,\mC)$ \\
We take $g_1 = e^{a_1X_1}$ and $g_2=e^{a_2 X_2}$. The coordinates are
\bea \begin{array}{llll}
x_1  =  r (\cos c - \frac{1}{2}(a_1^2+a_2^2) e^{ic}) &&& x_3  =  r a_2 e^{ic} \\
x_2  =  r a_1 e^{ic} &&& x_4  =  r(\sin c - \frac{i}{2}(a_1^2+a_2^2) e^{ic}) 
\end{array}  \label{eq:4C4} 
\eea
and we have
\beq
R_2=X_2, \qquad R_1=X_1.
\eeq
The operator $\Box_{4C}$ is again as in eq. (\ref{eq:359}) with
\beq
\bigtriangleup_{LB}(4)=\frac{\partial^2}{\partial c^2} + 2i \frac{\partial}{\partial c}
+e^{-2ic} \left( \frac{\partial^2}{\partial a_1^2}+\frac{\partial^2}{\partial a_2^2} \right). 
\eeq
Both variables ${a_1,a_2}$ are ignorable. However, ${X_1,X_2}$ is only a MASA 
of $o(4,\mC)$ not of $e(4,\mC)$.

\end{zoznamnum}

\subsection{A graphical formalism}

All separable subgroup type coordinates on $M(4,\mC)$ can be summarized using  subgroup 
diagrams similar to those of the real groups $O(n)$ and $O(n,1)$ \cite{PogIzW,PogW,VilSm,frisw}.
They are directly related to ``tree'' diagrams introduced in Ref. \cite{vilenkin,VKS}
and discussed in Ref. \cite{PogIzW,PogW}.

On Figs.1 and 2 we use rectangles to denote Euclidean groups $E(n,\mC)$ and circles to denote
$O(n,\mC)$. In both cases the value of $n$ is indicated inside the rectangle, or circle. 
Trapezoids
are used to denote maximal Abelian subgroups. Inside the trapezoid we indicate which
MASA is involved. A rectangle with $n=1$ denotes a one-dimensional unipotent subgroup. 
The corresponding algebra is represented by a nilpotent matrix.

Figs.(1a), \ldots, (1f) correspond to the coordinate systems 
(\ref{eq:V4100}), (\ref{eq:V4200}), (\ref{eq:V431}), (\ref{eq:V441}), (\ref{eq:V451}) and
 (\ref{eq:V462}). 
Similarly Figs.(2a), \ldots, (2d)
correspond to the coordinate systems (\ref{eq:4C1}), (\ref{eq:4C2}), (\ref{eq:4C3}) 
and (\ref{eq:4C4}).

\begin{figure}[]
\begin{center}
\noindent\includegraphics[height=3.6in]{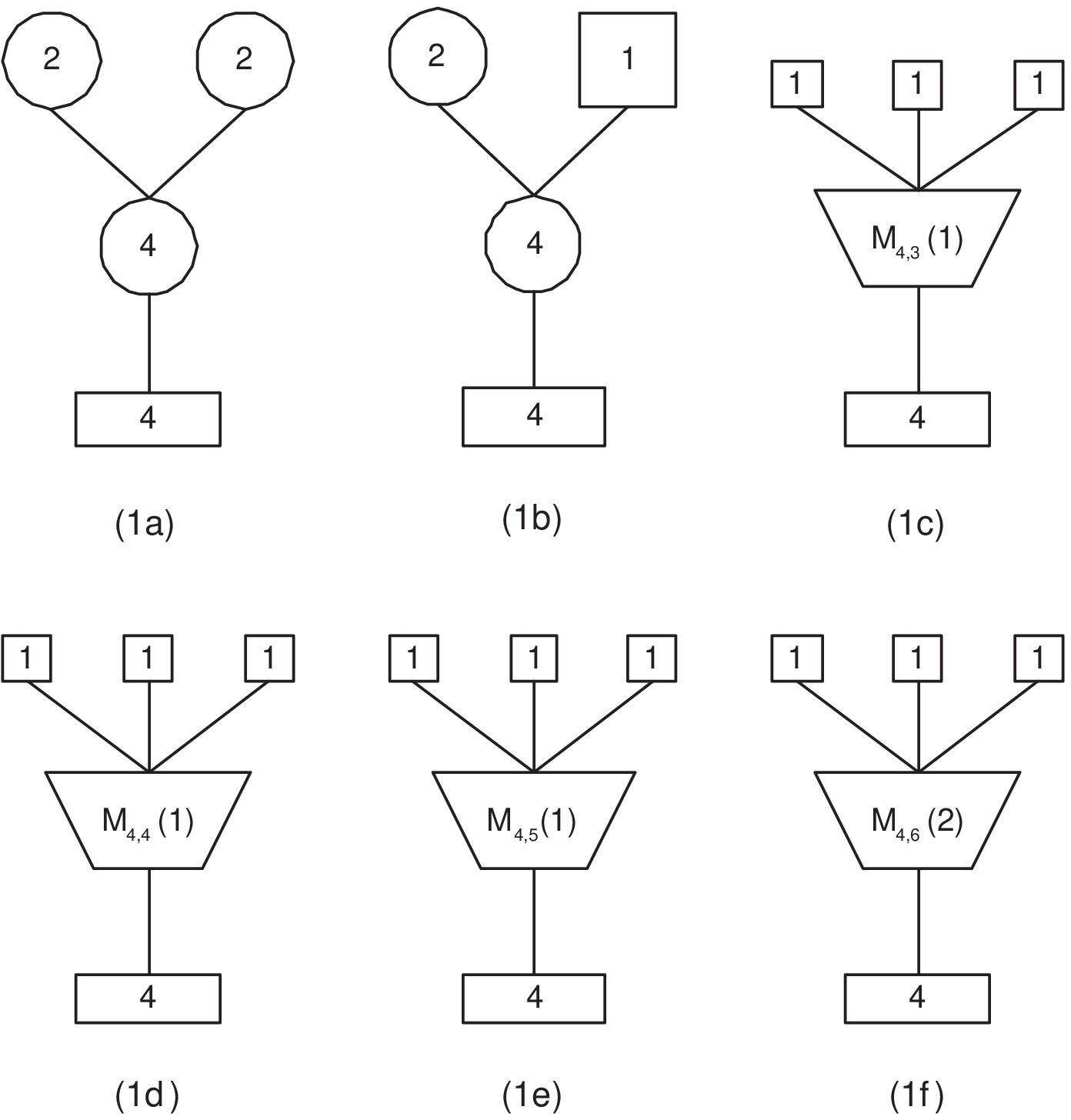}
\caption{Subgroup chains for $E(4,\mC)$ ending in maximal Abelian subgroups.}
\label{Fig.1}
\end{center}
\end{figure}

\begin{figure}[]
\begin{center}
\noindent\includegraphics[height=1.8in]{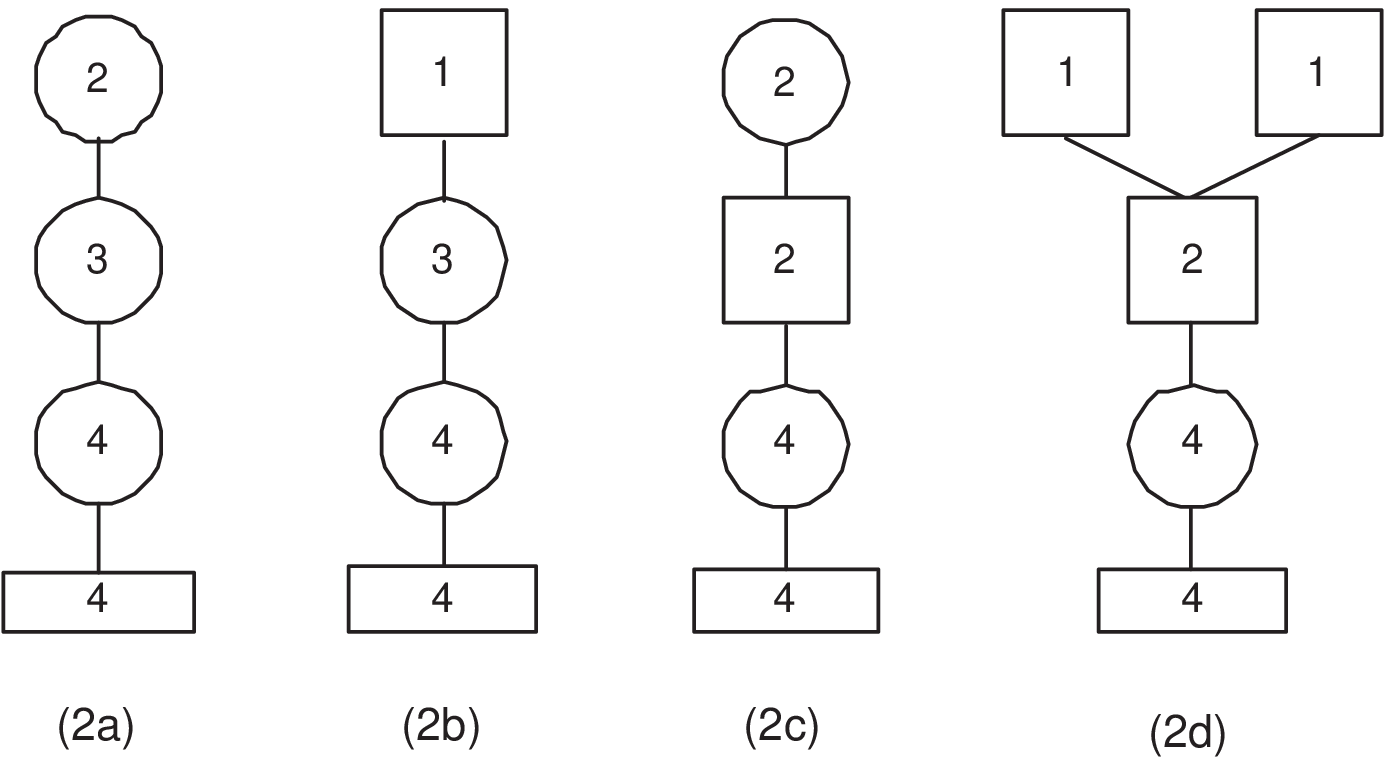}
\caption{Subgroup chains for $E(4,\mC)$ ending in nonmaximal Abelian subgroups.}
\label{Fig.2}
\end{center}
\end{figure}



\section{Subgroup type coordinates in the real spaces $M(4)$ and $M(3,1)$}
\setcounter{equation}{0}

When one passes from the complex Euclidean space $M(n,\mC)$ to the real 
Euclidean or pseudo-Euclidean spaces $M(p,q)$, $p \geq q \geq 0$, two 
different phenomena must be taken into account. Firstly some subgroups, in particular 
Abelian ones, that exist for $E(n,\mC)$ may have counterparts only for certain 
signatures $(p,q)$. Secondly in real spaces with $q \geq 1$ vectors can have positive, 
negative and zero length. 
The existence of two types of anisotropic vectors leads to a proliferation of subgroup 
chains and of coordinate systems.

\subsection{The real Euclidean space $M(4)$}

Only two of the subgroup chains illustrated on Fig.1 and Fig.2 are realized in this case, 
namely those of Fig.(1a) and Fig.(2a). The corresponding systems of separable coordinates are 
(\ref{eq:V4100}) with 
$0 \leq r \leq \infty$ , $0 \leq c \leq \frac{\pi}{2}$, $0 \leq a < 2 \pi$, 
$0 \leq b < 2 \pi$ 
and (\ref{eq:4C1}) with 
$0 \leq r \leq \infty, \, 0 \leq c \leq \pi,\, 0 \leq b \leq  \pi ,\,  0 \leq a < 2 \pi$.

All other subgroup chains lead to decomposable coordinate systems. The decomposition 
patterns are $4=3+1$, $4=2+2$, $4= 2+1+1$ and $4=1+1+1+1$.

\subsection {The real Minkowski space $M(3,1)$}

This case is somewhat richer than the previous one. Indeed a subgroup 
$O(n,\mC) \subset E(n,\mC)$ can correspond to $O(n) \subset E(n)$ or $O(n-1,1) \subset E(n)$
in the real case. We denote $O(n)$ by a circle, $O(n-1,1)$ by a semicircle (``hyperbola'')
on the corresponding subgroup diagrams.

Among the six coordinates of subgroup type in $M(4,\mC)$, related to MASAs, only three 
are represented on $M(3,1)$, see Fig.3.

The Cartan subalgebra (\ref{eq:MV41}) goes into $o(2) \oplus (1,1) \subset o(3,1)$ 
and the corresponding cylindrical coordinates (\ref{eq:V4100}) go into
\bea \begin{array}{rcll}
x_1 & = & r \sinh c \cos a   & \qquad 0 \leq a < 2 \pi  \\
x_2 & = & r \sinh c \sin a & \qquad 0 \leq b < \infty  \\
x_3 & = & r \cosh c \sinh b & \qquad 0 \leq c < \infty \\
x_4 & = & r \cosh c \cosh b.   \end{array}
\eea

The MASA $M_{4,2}(0)$ has no analogue in $o(3,1)$. Among the $k_0=1$ MASAs 
$M_{4,3}(1)$ and $M_{4,4}(1)$ have analogs in $M(3,1)$ while $M_{4,5}(1)$ does not. 
Finally $k_0=2$ is not allowed. The coordinate system corresponding to $M_{4,3}(1)$ 
in the space $M(3,1)$ is as in (\ref{eq:V431}) with 
$0 \leq r \leq \infty$, $ - \infty < a_i < \infty$ ($i=1,2$), $- \infty < z < \infty$.
For $M_{4,4}(1)$ the coordinates are as in (\ref{eq:V441}) with $ -1 \leq \beta \leq 1$ 
and $r,a_1,a_2,z$ as for the case of $M_{4,3}(1)$. Thus the only three subgroup diagrams of 
Fig.1 that give rise to subgroup diagrams and separable coordinates on $M(3,1)$, 
are (1a)$\rightarrow$(3a), (1c)$\rightarrow$(3b), (1d)$\rightarrow$(3c).

The four subgroup chains of Fig.2 give rise to the 6 chains of Fig.4. More specifically 
we have (2a)$\rightarrow$(4a),(4b) and (4c), (2b)$\rightarrow$(4d),
(2c)$\rightarrow$(4e) and (2d)$\rightarrow$(4f). We do not detail the 6 different types of 
coordinates systems on the upper sheet of the two sheeted hyperboloid 
$x_1^2+x_2^2+x_3^2-x_0^2=-1$ here. They can be found in e.g. \cite{VilSm,frisw} 
and \cite{PogW}.

\begin{figure}[]
\begin{center}
\noindent\includegraphics[height=1.8in]{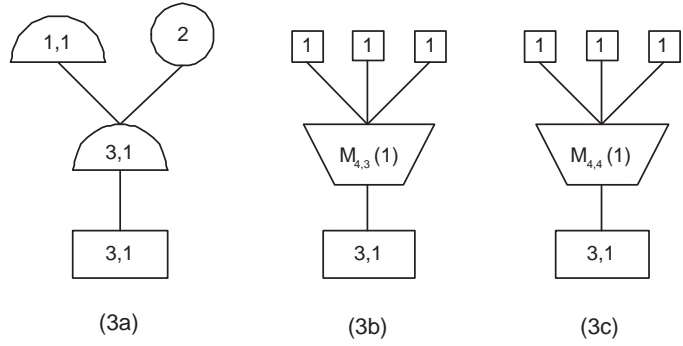}
\caption{Subgroup chains for $E(3,1)$ ending in maximal Abelian subgroups.}
\label{Fig.3}
\end{center}
\end{figure}

\begin{figure}[]
\begin{center}
\noindent\includegraphics[height=1.8in]{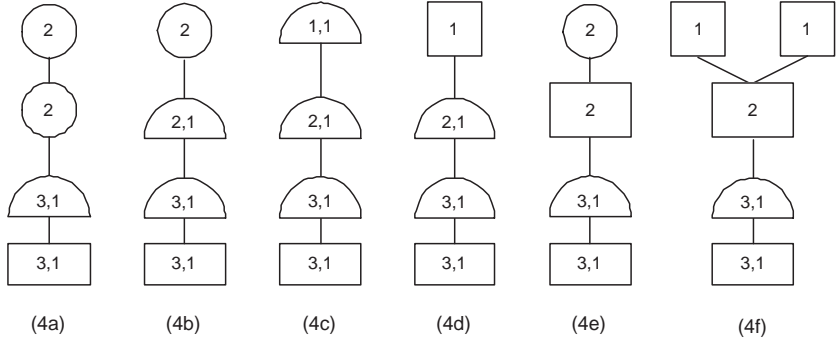}\caption{Subgroup chains for $E(3,1)$ ending in nonmaximal Abelian subgroups.}
\label{Fig.4}
\end{center}
\end{figure}



\section{Subgroup type coordinates in the real space $M(2,2)$}
\setcounter{equation}{0}

\subsection{Structure of MASAs of $e(2,2)$}.

A complete classification of MASAs of $e(p,2)$, in particular $e(2,2)$, was 
performed earlier \cite{epq}. It is somewhat more complicated than that of $e(4,\mC)$ 
although similar. A MASA of $e(p,q)$ contains $k=\kz+\kp+\km$ mutually orthogonal 
translations, where $\kz$, $\kp$ and $\km$ are the numbers of isotropic, positive 
length and negative length translations, respectively. For $p \geq q$ we have
\beq
0\leq \kz \leq q, \quad 0 \leq \kp \leq p, \quad 0 \leq \km \leq q, \quad 0 \leq \kz+\kp+\km \leq p+q.
\eeq
We are mainly interested in MASAs with $\kp=\km=0$, since the presence of any positive
or negative length translations leads to a decomposition of the space $M(p,q)$
and to decomposable coordinate systems.

We  proceed as in the case of $e(4, \mC)$. The algebra $e(2,2)$ is realized by 
$5 \times 5$ matrices as in eq. (\ref{eq:E}) and the metric $K=K^T \in \mR^{4 \times 4}$  has 
the signature $(2,2)$.

As in the case of $e(4,\mC)$ we consider $\kz=0,1$ and $2$ separately. Each subgroup chain 
and the corresponding coordinate system on $M(4, \mC)$ gives rise to at least
one system on $M(2,2)$.

\subsection{MASAs of $e(2,2)$ with $\kz=0$}

We are dealing here with MASAs of $o(2,2)$ that are also maximal in $e(2,2)$. The
algebra $o(2,2)$ has three inequivalent Cartan subalgebras and so 3 different coordinate
systems correspond to the cylindrical coordinates (\ref{eq:V4100}). In each case we  
give the subgroup chain, the group action on the origin, the coordinates and the complete 
set of commuting operators.

\begin{zoznamnum}
\item
The compact Cartan subalgebra $o(2) \oplus o(2)$. \\
The subgroup chain is 
\beq
E(2,2) \supset O(2,2) \supset O(2) \otimes O(2).
\eeq
We use the diagonal metric $K={\rm diag}(1,1,-1,-1)$ and put
\beq
| y \rangle = e^{-a(E_{12}-E_{21})} e^{-b(E_{34}-E_{43})} e^{c(E_{13}+E_{31})}e^{rE_{15}}
 |0 \rangle.
\eeq
The coordinates are
\bea \begin{array}{rcll}
x_1 & = & r \cosh c \cos a   & \qquad 0 \leq r < \infty  \\
x_2 & = & r \cosh c \sin a & \qquad 0 \leq a < 2 \pi, \qquad 0 \leq b < 2 \pi  \\
x_3 & = & r \sinh c \cos b & \qquad 0 \leq c < \infty \\
x_4 & = & r \sinh c \sin b.   \end{array} \label{eq:54_car}
\eea
The Laplace operator and complete set of commuting operators are
\bea 
\Box_{2,2}&=&\frac{\partial^2}{\partial r^2} + \frac{3}{r} \frac {\partial}{\partial r} 
+\frac {1}{r^2} \bigtriangleup_{LB}(2,2) \label{eq:55}\\
\bigtriangleup_{LB}(2,2)& =&  -\frac{\partial^2}{\partial c^2} 
-\frac{2(e^{4c}+1)}{e^{4c}-1}\frac{\partial}{\partial c}                                     
+\frac{4e^{2c}}{(e^{2c}+1)^2} \frac{\partial^2}{\partial a^2}
-\frac{4e^{2c}}{(e^{2c}-1)^2} \frac{\partial^2}{\partial b^2}  \nonumber 
\eea
\beq
\{\Box_{2,2},\bigtriangleup_{LB}(2,2), L_{12}=E_{12}-E_{21}, L_{34}=E_{34}-E_{43} \}.
\eeq

\item
The noncompact Cartan subalgebra $o(1,1) \oplus o(1,1)$. \\
The subgroup chain is 
\beq
E(2,2) \supset O(2,2) \supset O(1,1) \otimes O(1,1).
\eeq
Again in the diagonal metric $K={\rm diag}(1,1,-1,-1)$  we have 
\beq
| y \rangle = e^{a(E_{13}+E_{31})} e^{b(E_{24}+E_{42})} e^{c(E_{14}+E_{41})}e^{rE_{15}} |0 \rangle.
\eeq
The coordinates are
\bea \begin{array}{llll}
x_1  =  r \cosh c \cosh a  &&& x_3  =  r \cosh c \sinh a   \\
x_2  =  r \sinh c \sinh b &&& x_4  =  r \sinh c \cosh b. 
\end{array}  \label{eq:510_noncomp}
\eea  
The Laplace operator is as in eq. (\ref{eq:55}) with
\bea
\bigtriangleup_{LB}(2,2) = -\frac{\partial^2}{\partial c^2} 
-\frac{2(e^{4c}+1)}{e^{4c}-1}\frac{\partial}{\partial c}
-\frac{4e^{2c}}{(e^{2c}+1)^2} \frac{\partial^2}{\partial a^2}
+\frac{4e^{2c}}{(e^{2c}-1)^2} \frac{\partial^2}{\partial b^2} 
\eea
and the commuting set is
\beq
\{\Box_{2,2},\bigtriangleup_{LB}(2,2), L_{13}=E_{13}+E_{31}, L_{24}=E_{24}+E_{42} \}.
\eeq

\item
The Cartan subalgebra $o(2) \oplus o(1,1)$. \\
The algebra $o(2,2)$ has a third Cartan subalgebra, with one compact and one 
noncompact basis element. 
The subgroup chain is
\beq
E(2,2) \supset O(2,2) \supset O(2) \otimes O(1,1). \label{eq:513}
\eeq
We represent this cartan subalgebra by the matrices
\bea X =\left( \begin{array}{ccccc} a & -b  & 0 & 0 & 0 \\
                                    b &  a  & 0 & 0 & 0 \\
                                    0 &  0  & 0 & 0 & 0   \\
                                    0 &  0  & -a & -b & 0  \\
                                    0 &  0  &  b & -a & 0 
                  \end{array} \right), \quad
K=\left( \begin{array}{cc} 0 & I_2 \\
                           I_2 & 0 \\
         \end{array} \right), \label{eq:514}
\eea
where we are using a nondiagonal metric. As a MASA (\ref{eq:513}) is decomposable but
not absolutely orthogonally indecomposable (D but NAOID) \cite{verop,epq} (i.e., after 
complexification the matrix $X$ can be diagonalized).

The appropriate group action is
\bea \! \! \! \! \! \! \! \! \! \! \! \! \! \! \! \! \! \! \! \! \! \!
| y \rangle = e^{a(E_{11}+E_{22}-E_{33}-E_{44})} e^{b(-E_{21}+E_{12}-E_{34}+E_{43})} 
e^{c(E_{12}+E_{21}-E_{34}-E_{43})}e^{r(E_{15}+E_{35})/ \sqrt 2} |0 \rangle \nonumber \\
\eea
so we have 
\bea 
x_1 & = & \frac{r}{\sqrt 2} e^a (\cosh c \cos b - \sinh c \sin b)   \nonumber \\
x_2 & = & \frac{r}{\sqrt 2} e^a (\cosh c \sin b + \sinh c \cos b) \nonumber \\
x_3 & = & \frac{r}{\sqrt 2} e^{-a} (\cosh c \cos b + \sinh c \sin b) \label{eq:516}\\
x_4 & = & \frac{r}{\sqrt 2} e^{-a} (\cosh c \sin b - \sinh c \cos b). \nonumber 
\eea
The Laplace operator in these coordinates is as in eq. (\ref{eq:55}) with
\bea \! \! \! \! \! \! \! \! \! \! \! \! \! \! \! \! \! \! \! \!
\bigtriangleup_{LB}(2,2) =-\frac {\partial^2}{\partial c^2} 
- 2\tanh 2c \frac{\partial}{\partial c}  
+ 2\frac{\sinh 2c}{(\cosh 2c)^2} \frac{\partial^2}{\partial a \partial b} 
+ \frac{1}{(\cosh 2c)^2} \left( \frac {\partial^2}{\partial a^2}-\frac{\partial^2}{\partial b^2} \right). 
\label{eq:517}
\eea
The complete set of commuting operators is
\beq
\{\Box_{2,2},\bigtriangleup_{LB}(2,2), E_{11}+E_{22}-E_{33}-E_{44}, -E_{21}+E_{12}-E_{34}+E_{43} \}.
\eeq

\end{zoznamnum}

The relation between the real coordinates (\ref{eq:516}) and the complex cylindrical coordinates 
(\ref{eq:V4100}) is best seen if we transform (\ref{eq:MV41}), (\ref{eq:M41-00}) and 
(\ref{eq:V4100})
to the antidiagonal metric $K$ of (\ref{eq:514}). The transformation $gK_Dg^T=K$ and $gX_Dg^{-1}=X$,
where $K_D=I_4$, is realized by
\bea 
g= \frac{1}{2} \left( \begin{array}{cccc} 1 & i  & 1 & i \\
                                              -i & 1 & i & -1  \\
                                             1 & -i & 1 & -i     \\
                                             i & 1 & -i & -1 
                  \end{array} \right). \label{eq:519}
\eea
Putting $|x \rangle = g |x_D \rangle$ with $|y_D \rangle$ as in eq. (\ref{eq:V4100}) we obtain
\bea 
x_1 & = & \frac{1}{2} r e^{i\frac{\tilde{a}+\tilde{b}}{2}} \left( (\cos \tilde{c} +\sin \tilde{c})
\cos \frac{\tilde{a} -\tilde{b}}{2}+i((\cos \tilde{c} -\sin \tilde{c})
\sin \frac{\tilde{a} -\tilde{b}}{2}) \right)   \nonumber \\
x_2 & = & \frac{1}{2} r  e^{i\frac{\tilde{a}+\tilde{b}}{2}} \left( (\cos \tilde{c} +\sin \tilde{c})
\sin \frac{\tilde{a} -\tilde{b}}{2}-i((\cos \tilde{c} -\sin \tilde{c})
\cos \frac{\tilde{a} -\tilde{b}}{2}) \right) \nonumber \\
x_3 & = & \frac{1}{2} r e^{-i\frac{\tilde{a}+\tilde{b}}{2}} \left( (\cos \tilde{c} +\sin \tilde{c})
\cos \frac{\tilde{a} -\tilde{b}}{2}-i((\cos \tilde{c} -\sin \tilde{c})
\sin \frac{\tilde{a} -\tilde{b}}{2}) \right) \label{eq:520}\\
x_4 & = & \frac{1}{2} r e^{-i\frac{\tilde{a}+\tilde{b}}{2}} \left( (\cos \tilde{c} +\sin \tilde{c})
\sin \frac{\tilde{a} -\tilde{b}}{2}+i((\cos \tilde{c} -\sin \tilde{c})
\cos \frac{\tilde{a} -\tilde{b}}{2}) \right). \nonumber 
\eea
Putting
\bea
i \frac{\tilde{a}+\tilde{b}}{2}=a, \quad \frac{\tilde{a}-\tilde{b}}{2}=b, \quad 
\cos \tilde{c} + \sin \tilde{c} = \sqrt 2 \cosh c, \quad 
\cos \tilde{c} - \sin \tilde{c} = i \sqrt 2 \sinh c \nonumber
\eea
and restricting to $a,b,c \in \mR$, we obtain the coordinates (\ref{eq:516}).

The complex algebra $e(4, \mC)$ has only one further MASA with $\kz=0$, namely that 
of eq. (\ref{eq:M420}). It, however, has two distinct real forms. One of them, $M_1(0)$,
coincides with that of eq.(\ref{eq:M420}) with $a$ and $b \in \mR$. In $e(2,2)$ this MASA
is absolutely orthogonally indecomposable, but decomposable (AOID but D). The subgroup chain is
\beq
E(2,2) \supset O(2,2) \supset {\rm exp}(M_1(0)).
\eeq
The coordinates, the Laplace operator  and commuting  operators are as in (\ref{eq:V4200}),
(\ref{eq:316}) and (\ref{eq:317}) with all entries real.

The other MASA of $e(2,2)$, $M_2(0)$, is absolutely orthogonally indecomposable, indecomposable, 
but not absolutely indecomposable (AOID, ID but NAID). We represent $M_2(0)$ by the matrices:
\bea M_2(0) =\left( \begin{array}{ccccc} 0 & a  & 0 & b &0\\
                                        -a & 0 & -b & 0 & 0 \\
                                   0 & 0 & 0 & a & 0   \\
                                         0 & 0 & -a & 0 & 0  \\
                                    0 & 0 & 0 & 0 & 0 
                  \end{array} \right), \qquad \qquad
K =\left( \begin{array}{cc} 0 & I_2 \\
                                        I_2 & 0  \\
                                      \end{array} \right) \label{eq:M2_0}
\eea
The subgroup chain is
\beq
E(2,2) \supset O(2,2) \supset {\rm exp} M_2(0).
\eeq

The MASA $M_{4,2}(0) \subset e(4)$ (see eq. (\ref{eq:M420})) can be transformed into the form 
of (\ref{eq:M2_0}) by a similarity transformation with
\bea G= \frac{1}{\sqrt 2} \left( \begin{array}{cccc} 	1 & 0  & 0 & 1 \\
                                              		i & 0 & 0 & -i  \\
                                             		0 & 1 & 1 & 0     \\
                                         		0 & i & -i & 0 
                  \end{array} \right) \label{eq:525}
\eea
which preserves the metric $K$. To obtain the real space $M(2,2)$ we put 
\beq
a \rightarrow -ia, \qquad b \rightarrow ib, \qquad a,b \in \mR.
\eeq
The corresponding coordinates in $M(2,2)$ are obtained from (\ref{eq:V4200}) by the transformation
$|x \rangle \rightarrow G | x \rangle$ and we obtain 
\bea \begin{array}{lll}
x_1=\frac{1}{\sqrt2} r(e^{c} \cos a + b e^{-c}  \sin a)&&x_3=\frac{1}{\sqrt2} r e^{-c} \cos a \\
x_2=\frac{1}{\sqrt2} r(e^{c} \sin a - b e^{-c}  \cos a)&&x_4=\frac{1}{\sqrt2} r e^{-c} \sin a
\end{array} \label{eq:527M20}
\eea
with $0\leq r < \infty, 0 \leq a < 2 \pi, - \infty < c < \infty$ and $ 0 \leq b < \infty$.
These coordinates correspond to the group action
\beq
e^{a(-E_{12}+E_{21}-E_{34}+E_{34})}e^{b(E_{14}-E_{23})}
e^{c(-E_{11}-E_{22}+E_{33}+E_{44})}
e^{r(E_{15}+E_{35})}.
\eeq
The Laplace operator is 
\bea
\Box_{2,2}= \frac{\partial^2}{\partial r^2}
+\frac{3}{r} \frac{\partial}{\partial r}
+ \frac{1}{r^2}
\left( -\frac{\partial^2}{\partial c^2} + \frac{\partial}{\partial c} 
-4 e^{4c} \frac{\partial^2}{\partial b^2}
-4 e^{2c}\frac{\partial^2}{\partial a \partial b} \right).
\eea 
This corresponds to the case (\ref{eq:318}) plus (\ref{eq:317}) for $E(4,\mC)$.
The commuting operators are 
\beq
\{\Box_{2,2},\bigtriangleup_{LB}(2,2), E_{12}-E_{21}+E_{34}-E_{43}, E_{14}-E_{23} \}.
\eeq

\subsection{MASAs of $e(2,2)$ with $\kz=1$}

The maximal Abelian subalgebras of $e(2,2)$ with $\kz=1$ can be written exactly 
in the same form as those of $e(4, \mC)$ (see Section 3.3). All parameters and variables 
are real. As for $e(4, \mC)$, there are three different cases to consider. 
All can be written as in eq. (\ref{eq:313}). In eq. (\ref{eq:320}) the parameter 
$\beta$ is real and satisfies $-1 \leq \beta \leq 1$.

The three separable coordinate systems are (\ref{eq:V431}), (\ref{eq:V441}) and 
(\ref{eq:V451}). The Laplace operators and commuting operators are the same as 
in the complex case with all variables restricted to being real.

\subsection{MASAs of $e(2,2)$ with $\kz=2$}

For $\kz=2$ the MASAs of $e(2,2)$ have the same form as those of $e(4,\mC)$. There
are just two of them. The first leads to the coordinates (\ref{eq:V462}) with
$a_1,a_2,z$ and $r$ real. The Laplace operator is (\ref{eq:L337}).
The second MASA has the form (\ref{eq:338}) with real entries and, like (\ref{eq:338}),
does not provide a coordinate system.

\subsection{Decompositions of the space $M(2,2)$}

In view of the existence of a signature in $M(p,q)$ spaces, the four decompositions of 
$M(4,\mC)$ of eq. (\ref{eq:342})  give rise to six inequivalent decompositions of 
$M(2,2)$. 
They are:
\begin{zoznamnum}
\item $M(2,2) = M(2,1) \oplus M(0,1)$, or $M(1,2) \oplus M(1,0)$
\item $M(2,2) = M(2,0) \oplus M(0,2)$
\item $M(2,2) = M(1,1) \oplus M(1,1)$
\item $M(2,2) = M(2,0) \oplus 2M(0,1)$, or $M(0,2) \oplus 2M(1,0)$
\item $M(2,2) = M(1,1) \oplus M(1,0) \oplus M(0,1)$
\item $M(2,2) = 2M(1,0) \oplus 2M(0,1).$
\end{zoznamnum}
Each one-dimensional space corresponds to a Cartesian coordinate. The spaces
$M(2,0)$ and $M(0,2)$ correspond to polar coordinates (trigonometric ones),
the space $M(1,1)$ to ``hyperbolic polar coordinates", 
e.g. $x_1=\cosh \alpha$ and $x_2 = \sinh \alpha$. All subgroup type coordinates with 
a maximal number of ignorable variables on $M(2,1)$ (or $M(1,2)$) spaces were given 
earlier
\cite{conf}.

\subsection{Subgroup type coordinates on $M(2,2)$ with fewer ignorable variables}

Each of the subgroup chains in Section 3.6 has at least one analog for the group $E(2,2)$. 
Altogether that leads to five 
types of subgroup coordinates. Here we just give the subgroup chains and the corresponding coordinates.
The Laplacians and complete sets of commuting operators are easy to calculate. In all case we take 
the metric as $K={\rm diag}(1,1,-1,-1)$.
\begin{zoznamrom}
\item $E(2,2) \supset O(2,2) \supset O(2,1) \supset O(2)$
\bea \begin{array}{llll}
x_1=r \cosh c \cosh b \cos a &&& x_3=r \cosh c \sinh b \\
x_2 = r \cosh c \cosh b  \sin a &&& x_4 =r \sinh c.
\end{array} \label{eq:531}
\eea  
\item $E(2,2) \supset O(2,2) \supset O(2,1) \supset O(1,1)$
\bea \begin{array}{llll}
x_1 =r \cosh c \cos b \cosh a &&& x_3=r \cosh c \cos b \sinh a  \\
x_2 = r \cosh c \sin b &&& x_4 = r \sinh c. 
\end{array}\label{eq:532} 
\eea
\item $E(2,2) \supset O(2,2) \supset O(2,1) \supset E(1)$
\bea \begin{array}{llll}
x_1  =  r \cosh c (\cosh b -\frac{1}{2} a^2 e^b) &&& x_3=r \cosh c (\sinh b + \frac{1}{2} a^2 e^b)  \\
x_2 = r \cosh c \, a \, e^b &&& x_4= r \sinh c.
\end{array} \label{eq:533}
\eea
\item $E(2,2) \supset O(2,2) \supset E(1,1) \supset O(1,1)$
\bea \begin{array}{llll}
x_1 = r (\cosh c  -\frac{1}{2} b^2 e^c) &&& x_3 = r b e^c \sinh a   \\
x_2 =r b e^c \cosh a &&& x_4  =r (\sinh c + \frac{1}{2} b^2 e^c)  
\end{array} \label{eq:534}
\eea
\item $E(2,2) \supset O(2,2) \supset E(1,1) \supset E(1) \otimes E(1)$
\bea \begin{array}{llll}
x_1 = r (\cosh c  -\frac{1}{2} (a^2-b^2) e^c) &&& x_3 = r b e^c  \\
x_2 = r a e^c &&& x_4 = r (\sinh c + \frac{1}{2}(a^2-b^2) e^c).
\end{array} \label{eq:535} 
\eea
\end{zoznamrom}
All indecomposable coordinate systems on $M(2,2)$ and the corresponding subgroups 
chains are illustrated on Fig.5 and Fig.6.

We see that Fig.(1a), corresponding to the Cartan subalgebra of $O(4,\mC)$,
gives rise to Figs.(5a), (5b) and (5c) and thus to the coordinate systems
(\ref{eq:54_car}), (\ref{eq:510_noncomp}) and (\ref{eq:516}), respectively.
Fig.(1b) gives rise to Figs.(5d) and (5e). The corresponding coordinates on the 
space $M(2,2)$ are (\ref{eq:V4200}) (with real entries) and (\ref{eq:527M20}).
Figs.(1c), (1d), (1e) and (1f) give rise to Figs. (5f), (5g), (5h) and (5i), respectively.
The coordinates on $M(2,2)$ are the same as on $M(4,\mC)$ with real entries,
i.e.  (\ref{eq:V431}), (\ref{eq:V441}), (\ref{eq:V451}) and (\ref{eq:V462}), respectively.
Similarly Fig.(2a) gives rise to Figs.(6a) and (6b) and to the coordinates (\ref{eq:531}) 
and (\ref{eq:532}), respectively. Figs.(2b), (2c) and (2d) go to Figs.(6c), (6d) and (6e) and 
coordinates are (\ref{eq:533}), (\ref{eq:534}) and (\ref{eq:535}), respectively.

\begin{figure}[]
\begin{center}
\noindent\includegraphics[height=3.6in]{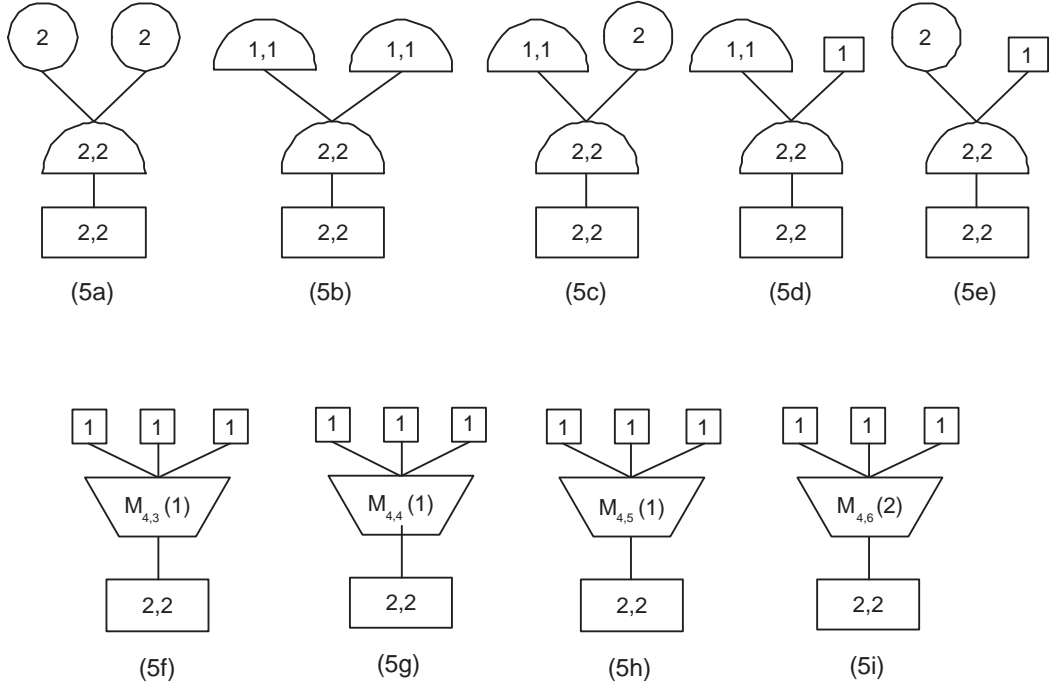}
\caption{Subgroup diagrams for $E(2,2)$ ending in maximal Abelian subgroups.}
\label{Fig.5}
\end{center}
\end{figure}

\begin{figure}[]
\begin{center}
\noindent\includegraphics[height=1.8in]{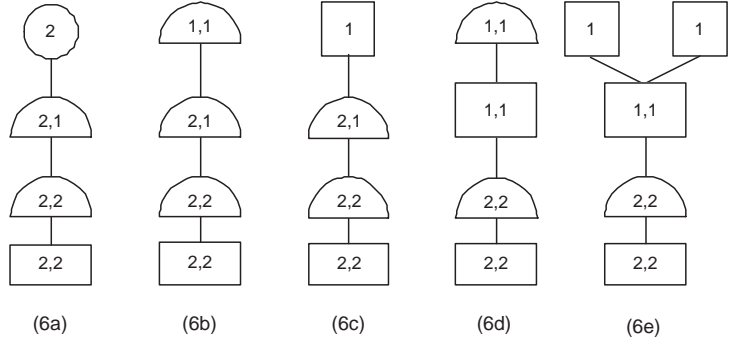}
\caption{Subgroup diagrams for $E(2,2)$ ending in nonmaximal Abelian subgroups.}
\label{Fig.6}
\end{center}
\end{figure}



\section{Solutions of the separated Schr\H{o}dinger equations in complex spaces}
\setcounter{equation}{0}

In this section we give the explicit solutions of the Schr\H{o}dinger equation
 (\ref{eq:schr})
 in the separable coordinate systems introduced above. We do this for the complex spaces 
$M(4,\mC)$ and $M(3,\mC)$. The results for the real spaces $M(n)$ \cite{morse} 
and $M(n,1)$ are known
 \cite{conf}.
Those for $M(2,2)$ can be obtained from the results of this section by imposing suitable 
reality conditions.

We  follow the diagrams on Figs.1 and 2.


\subsection{Subgroup chains involving link $E(4,\mC) \supset O(4,\mC)$}

The subgroup chains of Figs.(1a), (1b) and Figs.(2a), \ldots, (2d) all involve the subgroup link
$E(4,\mC) \supset O(4,\mC)$. The corresponding coordinate systems contain a four-dimensional 
(complex) 
radius $r$. This is a nonignorable variable. The coordinate systems corresponding to Figs.(1a),
(1b) and (2d) contain one more nonignorable variable denoted $c$. Those corresponding to Figs.(2a),
 (2b) and (2c) 
contain 3 nonignorable variables, $r$, $c$ and $b$.

We write the separated wave function for the cases illustrated on Figs.(1a), (1b) and (2d) as
\beq
\Psi(r,c,a,b)=R(r)C(c)e^{i(\alpha a+\beta b)}, \label{eq:61}
\eeq
where $\alpha$ and $\beta$ are (complex) constants of separation. The function $R(r)$
is always the same. It satisfies the equation
\beq
R''+\frac{3}{r} R'+\left( \frac{\lambda}{r^2}-E \right) R=0 
\label{eq:62}
\eeq
and is hence expressed in terms of cylindrical functions
\beq
R(r)=\frac{1}{r} J_{\nu}(\sqrt{-E}r) \, \, \, \, \, \, \, \nu=-5-\lambda. \label{eq:63}
\eeq

The function $C(c)$ in eq. (\ref{eq:61}) is different in each case.

For Fig.(1a), i.e. coordinates (\ref{eq:V4100}), we have 
\beq
C''+ 2 \cot 2c \,C'
- \left( \frac{\alpha^2}{\cos^2 c} + \frac{\beta^2}{\sin^2 c} + \lambda \right) C=0. \label{eq:64}
\eeq
Eq. (\ref{eq:64}) can be solved in terms of Jacobi functions as
\beq
C(c)=(\cos c)^\alpha (\sin c)^\beta P_n^{(\alpha,\beta)} (-\cos 2c), \, \, \, \, \, 
n=-\frac{1}{2}(\alpha+\beta+1\pm \sqrt{1-\lambda}).
\eeq
$P_n^{(\alpha,\beta)} (z)$ is a polynomial for $n \in Z^+_0$.

For Fig.(1b), i.e. coordinates (\ref{eq:V4200}), the equation for $C(c)$ is
\beq
C'' - 2 C' 
-(4e^{4c}\beta^2-4e^{2c}\alpha \beta +\lambda )C=0. \label{eq:67}
\eeq
Eq. (\ref{eq:67}) is solved in terms of Whittaker functions and we have
\beq
C(c)=W_{\frac{i \alpha}{2},\frac{1}{2} \sqrt{1-\lambda}}(2i \beta e^{2c}).
\eeq

For Fig.(2d), i.e. coordinates (\ref{eq:4C4}), the equation for $C(c)$ is 
\beq
C''+2iC'-(k^2 e^{-2ic}+\lambda)C=0 \label{eq:68}
\eeq
and is solved in terms of cylindrical functions
\beq
C(c)=e^{-ic}Z_{\nu}(k e^{-ic}), \, \, \, \, \nu=\pm\sqrt{1-\lambda}. \label{eq:69}
\eeq

Solutions corresponding to the subgroup chains on Figs.(2a), (2b) and (2c) will have the form
\beq
\Psi(r,c,b,a)=R(r)C(c)B(b)e^{i\alpha a}, \label{eq:610}
\eeq
where $\alpha$ is a constant of separation.
In all cases we have $R(r)$ as in eqs. (\ref{eq:62}) and (\ref{eq:63}).

On both Fig.(2a) and Fig.(2b) we have an $O(4, \mC) \supset O(3, \mC)$ link corresponding
to the variable $c$ (in eqs. (\ref{eq:4C1})   and (\ref{eq:4C2})). The function 
$C(c)$ in both cases satisfies
\beq
C'' - 2  \tan c \, C'
+ \left( \frac{k}{\cos^2 c} -\lambda \right) C=0 \label{eq:611}
\eeq
and hence
\beq
C(c)=\frac{1}{\cos c} P_\nu^\mu (i \tan c), \, \, \, \, \, \, \, 
\nu=\frac{1}{2}(-1 \pm \sqrt{1-4k}), \, \, \, \, \mu=\sqrt{1-\lambda}, \label{eq:612}
\eeq
where $P_\nu^\mu (z)$ is an associated Legendre function.

The function $B(b)$ is different in the two cases.

For Fig.(2a) we have an $O(3,\mC) \supset O(2,\mC)$ link and correspondingly
\beq
B'' -   \tan b \, B' 
-(\frac{\alpha^2}{\cos^2 b} +k )B=0 \label{eq:613}
\eeq
\beq
B(b)=\frac{1}{\sqrt{\cos b}} P_\nu^\mu (i \tan b), \, \, \, \, \, \, \, 
\nu=-\frac{1}{2} \pm \alpha, \, \, \, \, \mu=\sqrt{ \frac{1}{4}-k}. \label{eq:614}
\eeq

The corresponding subgroup link on Fig.(2b) is $O(3,\mC) \supset E(1,\mC)$ and we have
\beq
B'' + i B'
-(\alpha^2 e^{-2ib} +k)B=0  \label{eq:615}
\eeq
\beq
B(b)={\rm exp}(-\frac{1}{2}ib) Z_\nu (\pm \alpha e^{-ib}), \, \, \, \, \, \, \, 
\nu=\sqrt{\frac{1}{4}-k}, \label{eq:616}
\eeq 
where $Z_{\nu}(z)$ is a cylindrical function.

The subgroup chain on Fig.(2c) involves the link $O(4,\mC) \supset E(2,\mC)$ which  also figures on
Fig.(2d). Correspondingly the function $C(c)$ satisfies eq. (\ref{eq:68}) and is 
given by eq. (\ref{eq:69}). The final link on Fig.(2c) corresponds to $E(2, \mC) \supset O(2,\mC)$ 
and the function $B(b)$ satisfies
\beq
B'' +  \frac{1}{b}  B 
- (\frac{\alpha^2}{b^2} +k)B=0 \label{eq:617}
\eeq
\beq
B(b)= Z_\nu (\sqrt{-k} b), \, \, \, \, \, \, \, 
\nu=\pm \alpha, \label{eq:618}
\eeq
where $Z_{\nu}(z)$ is a cylindrical function.


\subsection{Subgroup chains leading directly to maximal Abelian subgroups of $E(4,\mC)$}

The  coordinate systems corresponding to Fig.(1c), \ldots, Fig.(1f) all have three 
ignorable variables and so we can write the corresponding wave functions as
\beq
\Psi= R(r)e^{\zeta z}e^{\alpha_1 a_1}e^{\alpha_2 a_2}, \label{eq:619}
\eeq
where $\zeta$, $\alpha_1$ and $\alpha_2$ are constants.

The coordinates are all nonorthogonal and the differential equations for $R(r)$ are all
linear first-order equations. The functions $R(r)$ are in all cases elementary ones. We do not 
give the equations satisfied by the functions $R(r)$, but only give the corresponding
 four solutions.

Fig.(1c) and coordinates (\ref{eq:V431}):
\beq
R(r)= \frac{1}{r} {\rm exp} \left( \frac{1}{2 \zeta} 
\left( \frac{\alpha_1^2+\alpha_2^2}{r} +Er \right) \right) . 
\label{eq:620}
\eeq

Fig.(1d) and coordinates (\ref{eq:V441}):
\beq
R(r)=((r+1)(r+\beta))^{- \frac{1}{2}}
{\rm exp} \left( \frac{1}{2 \zeta} 
\left( \frac{\alpha_1^2}{r+1}+\frac{\alpha_2^2}{r+\beta}+Er \right) \right).
 \label{eq:621}
\eeq

Fig.(1e) and coordinates (\ref{eq:V451}):
\beq
R(r)=\frac{1}{r+\kappa} {\rm exp} \left( \frac{1}{2 \zeta} \left( -\frac{\alpha_1^2}{(r+\kappa)^2}+
\frac{2 \alpha_1 \alpha_2}{(r+\kappa)}+Er \right) \right). \label{eq:622}
\eeq

Fig.(1f) and coordinates (\ref{eq:V462}):
\beq
R(r)={\rm exp} \left( \frac{1}{2 \alpha_1} \left( -\alpha_2^2r^2+(E-2\alpha_2 \zeta)r \right) \right). 
\label{eq:623}
\eeq


\subsection{Indecomposable coordinate systems in the space $M(3, \mC)$}

As we have seen in Section 3.5, two indecomposable separable coordinate systems of  subgroup type  
exist
in $M(3,\mC)$, namely (\ref{eqM3C_kapa0}) and (\ref{eqM3C_kapa1}).

In both cases we write the separated wave function as
\beq
\Psi(r,a,z)=R(r)e^{\alpha a+ \zeta z}. \label{eq:624}
\eeq
The two cases are quite different.

In coordinates (\ref{eqM3C_kapa0}) the equation for $R(r)$ is of first order:
\beq
2 \zeta R'+(\frac{1}{r^2} \alpha^2 +\frac{1}{r} \zeta -E )R=0 \label{eq:625}
\eeq
and its solution is
\beq
R(r)=\frac{1}{\sqrt r} {\rm exp} \left( \frac{1}{2 \zeta}(\frac{\alpha^2}{r}+Er) \right) . \label{eq:626}
\eeq

In the system (\ref{eqM3C_kapa1})  we obtain a second-order equation
for $R(r)$, namely
\beq
R'' +(2r \zeta^2 -2 \alpha \zeta -E)R=0. \label{eq:627}
\eeq
Equation (\ref{eq:627}) can be solved in terms of Airy functions (related 
to Bessel functions with index $\nu=1/3$)
\beq
R(r)=Ai(x), \, \, \, 
x= \zl r-\frac{\alpha}{\zeta}-\frac{E}{2 \zeta^2} \zr \zl 2 \zeta^2 \zr^\frac{1}{3}. 
 \label{eq:628}
\eeq



\section {Conclusions}
We have provided a complete classification of all subgroup type coordinates in
complex and real four-dimensional flat spaces in which the free Schr\H{o}dinger and
Hamilton-Jacobi equations allow  separation of variables. The results are best 
summed up in terms of the subgroup diagrams that are presented in this article.

For the complex Euclidean group $E(4,\mC)$ and the corresponding complex
space $M(4,\mC)$ there are six subgroup chains on Fig.1 leading to maximal Abelian
subgroups. In four cases these Abelian subgroups are 3-dimensional. The corresponding
maximal Abelian subalgebras, $M_{4,3}(1)$, $M_{4,4}(1)$, $M_{4,5}(1)$ and $M_{4,6}(2)$, each
provide three first-order operators in the Lie algebra $e(4,\mC)$ which together with
the Laplace operator $\Box_{4 \mC}$ form complete sets of commuting operators.
The subgroup chains in these four cases lead directly from $E(4,\mC)$ to a maximal
Abelian subgroup. Other subgroups could be included between $E(4,\mC)$
and the Abelian subgroups. However, they do not have second-order Casimir operators
and play no role in the problem of separating variables. The remaining two MASAs, 
the Cartan subalgebra $M_{4,1}(0)$ and the orthogonally indecomposable MASA $M_{4,2}(0)$, 
are two-dimensional. The missing operator in the complete set is provided by 
a Casimir operator of $O(4,\mC)$, i.e. the Laplace-Beltrami operator on
the complex sphere $S_4(\mC)$.

Four more subgroup diagrams are given on Fig.2. The corresponding subgroup chains
end in Abelian subgroups that are not maximal. In three cases they are just one-dimensional 
(Figs.(2a), (2b) and (2c)). Two missing commuting operators are provided by the Casimir operators 
of the intermediate groups in the chains. Fig.(2d) corresponds to a two-dimensional Abelian subgroup
at the end of the chain. It again is not a maximal Abelian subgroup.

For the real Euclidean group $E(4,\mR)$ and the corresponding Euclidean space $M(4,\mR)$
the situation is much simpler. Only two subgroups chains leading to indecomposable separable
coordinate systems exist, namely those of Fig.(1a) and Fig.(2a). The corresponding separable
coordinates are cylindrical
and spherical, respectively.

For the inhomogeneous Lorentz group $E(3,1)$ and the corresponding Minkowski
space $M(3,1)$ the situation is much richer and is illustrated on Fig.3 and Fig.4.

The full richness of the problem manifests itself in the case of a balanced signature, i.e. the
group $E(2,2)$ and the space $M(2,2)$. As shown on Fig.5, five subgroup chains lead through $O(2,2)$ 
to two-dimensional maximal Abelian subgroups. Those on Figs.(5a), (5b) and (5c) correspond to three different 
Cartan subalgebras. Figs.(5f), (5g), (5h) and (5i) correspond to different 3 dimensional maximal Abelian subgroups.
Finally the five diagrams of Fig.6 correspond to chains ending in Abelian subgroups
that are not maximal. The corresponding coordinate systems have less than the maximal
possible number of ignorable coordinates.

The results presented above for the complex space, $M(4,\mC)$, and the corresponding complex
Euclidean group, $E(4,\mC)$, are in complete agreement with the algebraic theory of
separation of variables in four-dimensional Riemannian spaces presented elsewhere 
\cite{BKM,KalMil1,KalMil2}.
Here we have added a group theoretical background to the theory. Moreover we have shown how coordinates
of subgroup 
type on a homogeneous space are directly generated by the action of the isometry group
$G$ of the space. To do this we needed a classification of the subgroups of $G$, in particular 
the Abelian ones.

More specifically for $M(4,\mC)$ the coordinates (\ref{eq:V4100}) corresponding to the Cartan subalgebra
are orthogonal. All other MASAs lead to nonorthogonal coordinates. The orthogonally indecomposable, but
decomposable MASA $M_{4,2}(0)$ contains both semisimple and nilpotent elements. 
It leads to a nonorthogonal separation; the ignorable coordinates $a$ and $b$  are the nonorthogonal 
ones (see eq.(\ref{eq:317})). 
The remaining MASAs, $M_{4,3}(1), \ldots, M_{4,6}(2)$, are all maximal Abelian nilpotent subalgebras
 (MANS). 
They all lead to nonorthogonal coordinate systems with three ignorable variables. Moreover 
the nonignorable 
variable $r$ is of ``first order''. Only first-order derivatives with respect to $r$ 
figure in all corresponding Laplace operators and
hence the separated solutions are expressed in terms of elementary functions.

All the coordinates of subgroup type of Section 3.6, involving nonmaximal Abelian subalgebras, 
are orthogonal.

In the case of the $M(2,2)$ space the situation is similar. Two of the Cartan subalgebras, 
namely the one isomorphic to $o(2) \oplus o(2)$ and the one isomorphic to $o(1,1) \oplus o(1,1)$,
are orthogonally decomposable and lead to orthogonal coordinates. The third Cartan subalgebra, 
isomorphic to $o(2) \oplus o(1,1)$  is decomposable, but 
orthogonally
indecomposable. It is, however, not absolutely orthogonally indecomposable.
It leads to nonorthogonal coordinates; the nonorthogonality concerns the ignorable variables 
only (see eq.(\ref{eq:517})). For all other coordinate systems the situation is the same 
as in the complex case.

The subgroup diagrams of Fig.1, \ldots , Fig.6 introduced in this article are
closely related to the tree diagrams used for compact groups \cite{V1,vilenkin,VKS} and
$O(n,1)$ \cite{PogW,frisw}. We have used the subgroup diagrams for $O(n,\mC)$
in Section 6 when constructing the separated wave functions in the different coordinate systems.
Very briefly the rules are as follows.
\begin{zoznamrom}
\item A link from a rectangle to a circle, i.e. $E(n,\mC) \supset O(n,\mC)$ corresponds to
a cylindrical function (as in eq. (\ref{eq:63})).
\item
A link from a circle to a rectangle, i.e. $O(n,\mC) \supset E(n-2,\mC)$ also leads to cylindrical 
functions (as in eqs. (\ref{eq:616}) and (\ref{eq:618})).
\item Links from circles to circles, i.e. $O(n, \mC) \supset O(n-1,\mC)$ or 
$O(n,\mC) \supset O(n_1,\mC) \otimes O(n_2, \mC)$ lead to Jacobi and Legendre functions.
\item
Diagrams involving trapezoids, i.e. subgroup chains including only Abelian subgroups lead to 
wave functions expressed in terms of elementary functions.
\end{zoznamrom}

These rules can be further elaborated and specified, but we leave this for a future study.

One of the possible applications of the present results is to study integrable and superintegrable 
systems in the spaces
considered. Indeed for each separable coordinate system in $M(4,\mC)$, $M(3,1)$, $M(2,2)$ 
or $M(4,\mR)$ we
can find a potential $V(x_1, \ldots, x_4)$ to be added to the kinetic energy term in
eq. (\ref{eq:schr})
such that the Schr\H{o}dinger
equation  still allows the separation of variables. The obtained Hamiltonians are integrable 
by construction:
the Hamiltonian commutes with three linearly independent differential 
second-order operators. The separable potentials will involve four
arbitrary functions of one variable each. Among the integrable and separable systems obtained
we can search for superintegrable and multiseparable systems. This, however,
goes beyond the scope of the present article.

A generalization of the present results to $M(n,\mC)$ and $M(p,q)$ spaces is under consideration. 
The tools are available since MASAs of $e(n,\mC)$ and $e(p,q)$ have already been studied 
\cite{epc,conf,epq}. Separable
coordinate systems for the real Euclidean space $M(n)$ are already known (not only the 
subgroup type ones)
 \cite{kalnins}.

\bigskip
\bigskip

\noindent
{\large \it Acknowledgments}

The research of E.G.K. was partly supported by the Marsden Fund. Z.T. thanks the 
Centre de recherches math{\'e}matiques for hospitality during her visits and SUNYIT for partial 
support of her research. The research of P.W. was partly supported by research grants from NSERC of Canada.


\bibliography{liter}
\bibliographystyle{plain}

\end {document}